\documentclass{IEEEtran}
\usepackage{amsmath,graphicx,amsthm,amsfonts}
\def\BibTeX{{\rm B\kern-.05em{\sc i\kern-.025em b}\kern-.08em
    T\kern-.1667em\lower.7ex\hbox{E}\kern-.125emX}}

\usepackage{romannum}
\usepackage{tikz}
\usetikzlibrary{dsp,chains,arrows,fit,spy}
\usepackage{caption}
\usepackage{enumitem}
\tikzset{
    font={\fontsize{9}{11.0476pt}\selectfont}}
\usepackage{mathtools,amssymb}
\usepackage{pgfplots}
\pgfplotsset{compat=newest}
\usepackage{stfloats}
\usepackage{multicol}
\usepackage{subcaption}
\DeclareMathOperator*{\argmin}{arg\,min}
\usepackage{algpseudocode}
\usepackage{algorithm}
\algnewcommand{\Parameters}[1]{%
    \State \textbf{\underline{Parameters}:}
    \Statex \hspace*{\algorithmicindent}\parbox[t]{.8\linewidth}{\raggedright #1}
}

\algnewcommand{\Input}[1]{%
    \State \textbf{\underline{Inputs}:}
    \Statex \hspace*{\algorithmicindent}\parbox[t]{.8\linewidth}{\raggedright #1}
}

\algnewcommand{\Initialization}[1]{%
    \State \textbf{\underline{Initialization}:}
    \Statex \hspace*{\algorithmicindent}\parbox[t]{.8\linewidth}{\raggedright #1}
}

\algnewcommand{\Iteration}[1]{%
    \State \textbf{\underline{Iteration}:}
    \Statex \hspace*{\algorithmicindent}\parbox[t]{.8\linewidth}{\raggedright #1}
}

\newtheorem{lemma}{Lemma}

\begin{document}

\title{\huge Study of Channel Estimation Algorithms for Large-Scale Multiple-Antenna Systems using 1-Bit ADCs and Oversampling}
\author{Zhichao~Shao, Lukas~T.~N.~Landau and Rodrigo~C.~de~Lamare}


\maketitle
\begin{abstract}
Large-scale multiple-antenna systems with large bandwidth are fundamental for future wireless communications, where the base station employs a large antenna array. In this scenario, one problem faced is the large energy consumption as the number of receive antennas scales up. Recently, low-resolution analog-to-digital converters (ADCs) have attracted much attention. Specifically, 1-bit ADCs are suitable for such systems due to their low cost and low energy consumption. This paper considers uplink large-scale multiple-antenna systems with 1-bit ADCs on each receive antenna. We investigate the benefits of using oversampling for channel estimation in terms of the mean square error and symbol error rate performance. In particular, low-resolution aware channel estimators are developed based on the Bussgang decomposition for 1-bit oversampled systems and analytical bounds on the mean square error are also investigated. Numerical results are provided to illustrate the performance of the proposed channel estimation algorithms and the derived theoretical bounds.
\end{abstract}




\section{Introduction}
Multi-user (MU) multiple-input multiple-output (MIMO) is currently
being used in many wireless communication systems like long-term
evolution (LTE), which allows for a small number of antennas at the
base station \cite{lte}. However, in the last decade the number of
wireless devices like mobiles, laptops and sensors, has experienced
an explosive growth and current MU-MIMO systems cannot serve such a
large number of users due to the limited bandwidth and increased
multi-user interference (MUI). With large antenna arrays at the base
station (BS), large-scale (or massive) MIMO can significantly
increase the spectral efficiency,  mitigate the propagation loss
caused by channel fading, reduce the MUI and have many other
advantages as compared to current systems \cite{Larsson,Rusek}. As
such, large-scale MIMO is a key technique for future wireless
communication systems, in which one favorable application is the
large-scale millimeter-wave (mmWave) communication system
\cite{Andrews}. However, many different configurations and
deployments need to be reconsidered. For example, by using current
high-resolution (8-12 bits) analog-to-digital converters (ADCs) for
each element of the antenna arrays at the BS, the hardware cost and
the energy consumption may become prohibitively high since the
dissipated power is exponentially scaled by the number of bits
\cite{Walden}.

The high cost and energy consumption associated with high-resolution ADCs has motivated the use of low-cost and low-resolution ADCs for large-scale MIMO systems. As one extreme case, 1-bit ADCs can largely reduce the hardware cost and energy consumption of the receiver. Many recent works have studied this area. For instance, the works in \cite{Jacobsson,Mo,Saxena,Landau,Usman,Li,Stockle,Shao,Shao2,Jeon} have studied massive MU-MIMO systems with coarsely quantized signals operating over frequency-flat, narrowband channels. The works in \cite{Jacobsson,Mo} have investigated the uplink channel capacity by MU-MIMO systems with 1-bit ADCs at the BS and \cite{Saxena,Landau,Usman} have analyzed different precoding techniques for the downlink. Regarding channel estimation, the studies in \cite{Li,Stockle,Shao} have proposed the Bussgang linear minimum mean squared error (BLMMSE), expectation-maximization (EM) based iterative hard thresholding (IHT) and recursive least squares (RLS) adaptive channel estimators, respectively. In the context of the signal detection used in uplink 1-bit massive MU-MIMO systems, the work in \cite{Shao2} proposes the iterative detection and decoding (IDD) technique together with regular LDPC codes and \cite{Jeon} presents a low-complexity near maximum-likelihood-detection (near-MLD) algorithm called 1-bit sphere decoding.

Moreover, some prior works have investigated 1-bit ADCs used in wideband communication systems. The works in \cite{Studer,Mollen,Mollen2,Jacobsson2} have studied massive MU-MIMO systems with coarsely quantized signals that deploy orthogonal frequency-division multiplexing (OFDM) for wideband communications. Their results show that it is satisfactory to use 1-bit ADCs in wideband massive MU-MIMO-OFDM systems. Furthermore, the studies in \cite{Zhang,Mo2,Mo3} have discussed some key transceiver design challenges, including channel estimation, signal detection, achievable rates and precoding techniques, in millimeter-Wave (mmWave) massive MIMO systems, which are promising candidates for 5G cellular systems.

The previous works have considered quantized systems with sampling at the Nyquist rate. However, utilizing oversampling at the receiver can partially compensate for the information loss brought by the coarse quantization \cite{Landau2}. The work in \cite{Ucuncu} has proposed faster than symbol rate (FTSR) sampling in an uplink massive MIMO system with coarsely quantized signals in terms of the symbol error rate (SER). It shows that the FTSR sampling provides about 5dB signal-to-noise ratio (SNR) advantage in terms of SER and achievable rate with a linear zero forcing receiver. The work in \cite{Landau3} has analyzed the achievable rate for 1-bit oversampled systems over band-limited channels. To reduce the computational cost caused by the large number of samples due to oversampling, a sliding window based linear detection has been proposed in \cite{Shao3}. In addition to the conventional system models based on matched filtering and correlated noise samples, alternative receiver assumptions exist in literature such as in \cite{Krone}, where the authors consider a wideband receiver whose bandwidth scales proportionally with the oversampling factor and has the drawback of additional received noise and interference from neighboring frequency bands.

From the channel estimation point of view, the works in \cite{Li,Stockle} have proposed different channel estimation techniques for systems operating at the Nyquist rate. However, only few works have considered channel estimation in oversampled systems. The study in \cite{Stein2} considers time-of-arrival estimation for systems with 1-bit quantization and oversampling and proposes corresponding performance bounds. The study in \cite{Schluter} has proposed carrier phase estimation and given lower bounds on complex channel parameter estimation for 1-bit oversampled systems based on \cite{Stein}. In the study in \cite{Ucuncu} the BLMMSE channel estimator is applied to the MIMO channel with 1-bit quantization and oversampling using the simplifying assumption of uncorrelated noise samples which then yields performance degradation especially at low SNR and high oversampling factors.

In this work, low-resolution aware (LRA) channel estimators are developed for 1-bit oversampled large-scale MIMO systems in the uplink based on the Bussgang decomposition. Although the received signals are quantized to 1 bit, the computations after the 1-bit ADCs of all algorithms compared are performed at a higher resolution (8 bits or higher). The application of oversampling at the receiver can lead to significantly better performance. Unlike prior works we explicitly consider the correlation of the filtered noise, which is a main property of oversampled systems, and employ the Bussgang decomposition \cite{Bussgang} to reformulate the nonlinear system into a statistically equivalent linear system. Based on this linear model, low-resolution aware least-squares (LS), linear minimum mean square error (LMMSE) and least-mean square (LMS) channel estimation algorithms are proposed for 1-bit oversampled systems and evaluate their computational costs. Moreover, an adaptive technique is devised to estimate the statistical quantities resulting from the Bussgang decomposition, which are required by channel estimators. We also examine the fundamental estimation limits by deriving a Bayesian framework and bounds on channel estimation for both non-oversampled and oversampled systems. In addition to the Bayesian Cram\'er-Rao bounds (CRBs), general CRBs is proposed for biased estimators due to the correlation between the signal and its quantization error. In summary, our work has the following contributions:

\begin{itemize}
    \item The LRA-LS, LRA-LMMSE and LRA-LMS channel estimation algorithms are presented for the 1-bit large-scale MIMO systems in the uplink with oversampling.
    \item We obtain analytical expressions associated with the Bayesian CRBs for the oversampled systems and observe that the proposed bounds are very close to the results obtained from simulations at low SNR.
    \item An adaptive technique is proposed to estimate the auto-correlation of the channel vector, which is an essential part for the Bussgang decomposition in 1-bit systems.
\end{itemize}

Some preliminary results have been shown in \cite{Shao4} and
\cite{Shao5}. However, as compared to \cite{Shao4,Shao5}, this paper
extends and refines the analysis of the correlation property of
filtered noise and proposes a more practical adaptive channel
estimator with lower computational cost. In the section of numerical
results, the performance of the proposed LRA-LMMSE estimator is
compared with its simplified version in \cite{Ucuncu}. Furthermore,
a comparison of the performance of systems using ADCs with more bits
is also shown in this paper.

The rest of this paper is organized as follows: Section \Romannum{2}
illustrates the system model and gives some statistical properties
of 1-bit quantization. Section \Romannum{3} derives the proposed
oversampling based channel estimators and analyzes the computational
complexity of the estimators. Section \Romannum{4} gives the upper
bounds of the Bayesian CRBs and the general CRBs for 1-bit
non-oversampled and oversampled MIMO systems. Section \Romannum{5}
compares the normalized mean square error (MSE) and SER performance
of the proposed and existing channel estimators. Section
\Romannum{6} concludes the paper.

Notation: The following notation is used throughout the paper.
Matrices are in bold capital letters and vectors in bold lowercase.
$\mathbf{I}_n$ denotes the $n\times n$ identity matrix and
$\mathbf{0}_n$ is the $n\times 1$ all-zero column vector.
Additionally, $\text{diag}(\mathbf{A})$ is a diagonal matrix only
containing the diagonal elements of $\mathbf{A}$. The transpose,
conjugate transpose and pseudoinverse of $\mathbf{A}$ are
represented by $\mathbf{A}^T$, $\mathbf{A}^H$ and $\mathbf{A}^+$,
respectively. $\mathbf{a}^*$ denotes the complex conjugate of
$\mathbf{a}$ and $[\mathbf{a}]_k$ represents the $k$th element of
vector $\mathbf{a}$. $(\cdot)^R$ and $(\cdot)^I$ get the real and
imaginary part from the corresponding vector or matrix,
respectively. $\otimes$ is the Kronecker product. Finally,
$\text{vec}(\mathbf{A})$ is the vectorized form of $\mathbf{A}$
obtained by stacking its columns and $\det(\mathbf{A})$ is the
determinant function. $\mathbf{x}\sim
\mathcal{CN}(\mathbf{a},\mathbf{B})$ indicates that $\mathbf{x}$ is
a complex Gaussian vector with mean $\mathbf{a}$ and covariance
matrix $\mathbf{B}$. The expectation and covariance is denoted as
$E\{\cdot\}$ and $Cov\{\cdot\}$, respectively.

\section{System Model and Problem Statement}

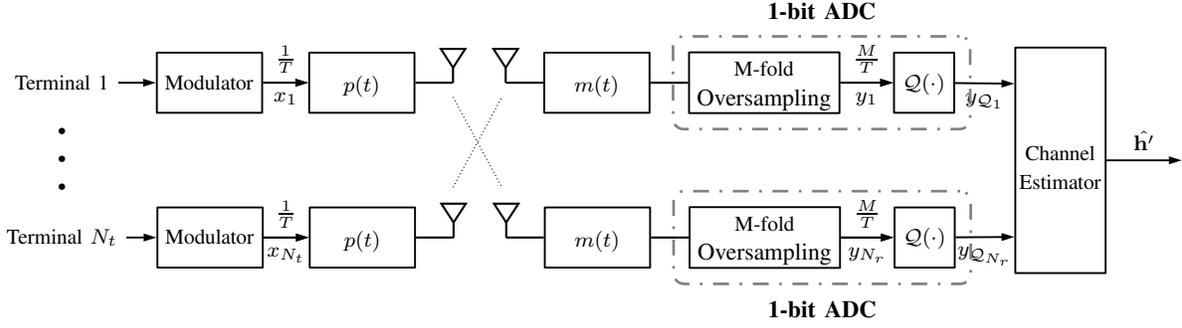
\begin{figure*}[!htbp]
    \centering
    \def\antenna{
	-- +(0mm,2.0mm) -- +(1.625mm,4.5mm) -- +(-1.625mm,4.5mm) -- +(0mm,2.0mm)
}
\tikzset{%
	harddecision/.style={draw, 
		path picture={
			\pgfpointdiff{\pgfpointanchor{path picture bounding box}{north east}}%
			{\pgfpointanchor{path picture bounding box}{south west}}
			\pgfgetlastxy\x\y
			\tikzset{x=\x*.4, y=\y*.4}
			%
			\draw (-0.5,-0.5)--(0,-0.5)--(0,0.5)--(0.5,0.5);  
			\draw (-0.25,0)--(0.25,0);
	}}
}

\begin{tikzpicture}
	\node (c0) {\footnotesize Terminal 1};
	\node[dspsquare, right= 0.5cm of c0,minimum width=1.4cm,text height=0.8em]       (c2) {\footnotesize Modulator};
	\node[dspsquare, right= 0.6cm of c2,minimum width=1.4cm]       (c10) {\footnotesize $p(t)$};
	\node[coordinate,right= 0.5cm of c10] (c3) {};

	\node[below= 0.25cm of c0] (c222) {\tiny \textbullet};
	\node[below= 0.04cm of c222] (c2222) {\tiny \textbullet};
	\node[below= 0.04cm of c2222] (c22222) {\tiny \textbullet};
	\node[coordinate,below= 0.62cm of c2] (cfix) {};
	
	\node[below= 1.6cm of c0] (c00) {\footnotesize Terminal $N_t$};
	\node[dspsquare, right= 0.4cm of c00,minimum width=1.4cm,text height=0.8em]                    (c22) {\footnotesize Modulator};
	\node[dspsquare, right= 0.6cm of c22,minimum width=1.4cm]       (c12) {\footnotesize $p(t)$};
	\node[coordinate,right= 0.5cm of c12] (c33) {};
	
	\node[coordinate,right= 0.7cm of c3] (c8) {};
	\node[coordinate,right= 0.7cm of c33] (c9) {};
	
	\node[coordinate,right= 0.5cm of c10.-15] (c88) {};
	\node[coordinate,below= 1.2cm of c88] (c99) {};
	
	\node[dspsquare, right= 0.5cm of c8,minimum width=1.4cm]       (c42) {\footnotesize $m(t)$};
	\node[dspsquare, right= 0.5cm of c9,minimum width=1.4cm]       (c43) {\footnotesize $m(t)$};
	
	\node[dspsquare, right= 0.5cm of c42,minimum width=2cm,text height=2em]       (c13) {\footnotesize M-fold \\ Oversampling};
	\node[dspsquare, right= 0.5cm of c43,minimum width=2cm,text height=2em]       (c14) {\footnotesize M-fold \\ Oversampling};
	
	\node[dspsquare,right= 0.7cm of c13,minimum width=0.8cm,text height=0.8em] (c15) {\footnotesize $\mathcal{Q}(\cdot)$};
	\node[dspsquare,right= 0.7cm of c14,minimum width=0.8cm,text height=0.8em] (c16) {\footnotesize $\mathcal{Q}(\cdot)$};
	
%
	
	\node[dspsquare,right= 10.7cm of cfix,minimum height=3cm,text height=2.5em,minimum width=1.2cm] (c17) {\footnotesize Channel \\\footnotesize Estimator };
	
	\node[coordinate,right= 1cm of c17] (c20) {};
	
	\node[coordinate,right= 0.65cm of c88] (c200) {};
	\node[coordinate,right= 0.65cm of c99] (c211) {};
	
	\draw[dspconn] (c0) -- node[] {} (c2);
	\draw[dspconn] (c2) -- node[midway,below] {\footnotesize $x_1$} (c10);
	\draw[dspconn] (c2) -- node[midway,above] {\footnotesize $\frac{1}{T}$} (c10);
	\draw[thick] (c10) -- node[] {} (c3);
	\draw[dspconn] (c00) -- node[] {} (c22);
	\draw[dspconn] (c22) -- node[midway,below] {\footnotesize $x_{N_t}$} (c12);
	\draw[dspconn] (c22) -- node[midway,above] {\footnotesize $\frac{1}{T}$} (c12);
	\draw[thick] (c12) -- node[] {} (c33);
	\draw [thick] (c3) \antenna;
	\draw [thick] (c33) \antenna;
	\draw [thick] (c8) \antenna;
	\draw [thick] (c9) \antenna;
	\draw[densely dotted] (c88) -- node[] {} (c211);
	\draw[densely dotted] (c99) -- node[] {} (c200);
	\draw[thick] (c8) -- node[] {} (c42);
	\draw[thick] (c9) -- node[] {} (c43);
	\draw[thick] (c42) -- node[midway,above] {} (c13);
	\draw[thick] (c43) -- node[midway,above] {} (c14);
	\draw[dspconn] (c13) -- node[midway,above] {\footnotesize $\frac{M}{T}$} (c15);
	\draw[dspconn] (c14) -- node[midway,above] {\footnotesize $\frac{M}{T}$} (c16);
	\draw[dspconn] (c13) -- node[midway,below] {\footnotesize $y_1$} (c15);
	\draw[dspconn] (c14) -- node[midway,below] {\footnotesize $y_{N_r}$} (c16);
	\draw[dspconn] (c15) -- node[midway,below] {\footnotesize $y_{\mathcal{Q}_1}$} (c17.121);
	\draw[dspconn] (c16) -- node[midway,below] {\footnotesize $y_{\mathcal{Q}_{N_r}}$} (c17.-121);
	\draw[dspconn] (c17) -- node[midway,above] {\footnotesize $\hat{\mathbf{h'}}$} (c20);

	\tikzset{blue dotted/.style={draw=black!50!white, line width=1pt,
			dash pattern=on 1pt off 4pt on 6pt off 4pt,
			inner sep=2mm, rectangle, rounded corners}};
	\node (first dotted box) [blue dotted, fit = (c13) (c15)] {};
	\node (second dotted box) [blue dotted, fit = (c14) (c16)] {};
	\node at (first dotted box.north) [above, inner sep=2mm] {\textbf{1-bit ADC}};
	\node at (second dotted box.south) [below, inner sep=2mm] {\textbf{1-bit ADC}};
	\end{tikzpicture}
    \caption{System model of 1-bit multi-user multiple-antenna system with oversampling at the receiver}
    \label{fig:system_model}
\end{figure*}

In this paper, we consider a single-cell multi-user large-scale MIMO
system with $N_t$ single-antenna terminals and a BS with $N_r$
receive antennas, where each receive antenna is equipped with two
1-bit ADCs (one for the in-phase component and the other for the
quadrature-phase component) and $N_r\gg N_t$. The system model is
depicted in Fig. \ref{fig:system_model}. In the uplink, by assuming
perfect synchronization the received oversampled signal
$\mathbf{y}\in\mathbb{C}^{MN_rN \times 1}$ can be expressed as
\begin{equation}
\mathbf{y}=\mathbf{H}\mathbf{x}+\mathbf{n},
\label{equ:system_model}
\end{equation}
where $\mathbf{x}\in\mathbb{C}^{NN_t \times 1}$ contains independent identically distributed (i.i.d.) transmitted symbols from $N_t$ terminals, each with block length $N$. The vector $\mathbf{x}$ is arranged as
\begin{equation}
\mathbf{x} = [x_{1,1} \quad \cdot\cdot\cdot \quad x_{N,1} \quad x_{1,2} \quad \cdot\cdot\cdot \quad x_{N,N_t}]^T,
\end{equation}
where $x_{i,j}$ corresponds to the transmitted symbol of terminal $j$ at time instant $i$. Each symbol has unit power so that $E[|x_{i,j}|^2]=1$. The vector $\mathbf{n}$ represents the filtered oversampled noise expressed by
\begin{equation}
\mathbf{n} = (\mathbf{I}_{N_r}\otimes \mathbf{G})\mathbf{w}
\label{equ_noise}
\end{equation}
with $\mathbf{w}\sim
\mathcal{CN}(\mathbf{0}_{3MN_rN},\sigma^2_n\mathbf{I}_{3MN_rN})$.
Note that the noise samples are described such that each entry of
$\mathbf{n}$ has the same statistical properties. Since in digital
domain the receive filter has a length of $2MN+1$ samples, $3MN$
unfiltered noise samples in the noise vector $\mathbf{w}$ need to be
considered for the description of an interval of $MN$ samples of the
filtered noise $\mathbf{n}$. The matrix $\mathbf{G}\in\mathbb{R}^{MN
\times 3MN}$ is a Toeplitz matrix that contains the coefficients of
the matched filter $m(t)$ (operated in analog domain) at different
time instants and is shown in (\ref{equ:R}), where
\begin{figure*}[!hb]
    \centering
    \hrulefill
    \begin{equation}
    \mathbf{G} = \begin{bmatrix}
    m(-NT)& m(-NT+\frac{1}{M}T)& \dots& m(NT) & 0 & \dots & 0\\
    0 & m(-NT)& \dots & m(NT-\frac{1}{M}T) & m(NT) & \dots & 0\\
    \vdots & \vdots & \ddots & \vdots & \vdots & \ddots & \vdots\\
    0 & 0 & \dots & m(-NT)& m(-NT+\frac{1}{M}T)& \dots& m(NT)\\
    \end{bmatrix}
    \label{equ:R}
    \end{equation}
\end{figure*}
$T$ is the symbol period and $M$ denotes the oversampling rate. The equivalent channel matrix $\mathbf{H}$ is described as
\begin{equation}
\mathbf{H} = [\mathbf{I}_{N_r}\otimes \mathbf{Z}(\mathbf{I}_{N}\otimes\mathbf{u})](\mathbf{H}'\otimes \mathbf{I}_N),
\label{equ:H}
\end{equation}
where $\mathbf{H}'\in\mathbb{C}^{N_r \times N_t}$ is the channel matrix for non-oversampled systems and $\mathbf{u}$ is an oversampling vector with length M, which has the form
\begin{equation}
\mathbf{u} = [0 \quad \cdots \quad 0 \quad 1]^T.
\end{equation}
The matrix $\mathbf{Z}\in\mathbb{R}^{MN \times MN}$ is a Toeplitz matrix that contains the coefficients of $z(t)$ at different time instants, where $z(t)$ is the convolution of the pulse shaping filter $p(t)$ and the matched filter $m(t)$ given by
\begin{equation}
\resizebox{.47\textwidth}{!}{$\displaystyle
    \mathbf{Z} = \begin{bmatrix}
    z(0) & z(\frac{T}{M}) & \dots & z(NT-\frac{1}{M}T)\\
    z(-\frac{T}{M}) & z(0) & \dots & z(NT-\frac{2}{M}T)\\
    \vdots & \vdots & \ddots & \vdots\\
    z(-NT+\frac{1}{M}T) & z(-NT+\frac{2}{M}T) & \dots & z(0)\\
    \end{bmatrix}.$}
\end{equation}
In particular, $M=1$ refers to the non-oversampling case.

Let $\mathcal{Q}(\cdot)$ represent the 1-bit quantization function, the resulting quantized signal $\mathbf{y}_\mathcal{Q}$ is given by
\begin{equation}
\mathbf{y}_\mathcal{Q}=\mathcal{Q}(\mathbf{y})=\mathcal{Q}(\mathbf{y}^R) + j\mathcal{Q}(\mathbf{y}^I).
\label{system_model}
\end{equation}
The real and imaginary parts of $\mathbf{y}$ are quantized
element-wised to $\{\pm\frac{1}{\sqrt{2}}\}$ based on the sign. The
factor $\frac{1}{\sqrt{2}}$ is to make the power of each quantized
signal to be one.

Since quantization strongly changes the properties of signals, some
statistical properties of quantization for Gaussian input signals
will be shown. For 1-bit quantization and Gaussian inputs, the
cross-correlation between the unquantized signal $\mathbf{s}$ with
covariance matrix $\mathbf{C}_\mathbf{s}$ and its 1-bit quantized
signal $\mathbf{s}_\mathcal{Q}$ is described by \cite{Bussgang}
\begin{equation}
\mathbf{C}_{\mathbf{s}_\mathcal{Q}\mathbf{s}}=\sqrt{\frac{2}{\pi}}\mathbf{K}\mathbf{C}_{\mathbf{s}},\mbox{where } \mathbf{K}=\text{diag}(\mathbf{C}_{\mathbf{s}})^{-\frac{1}{2}}.
\end{equation}
Furthermore, the covariance matrix of the 1-bit quantized signal $\mathbf{s}_\mathcal{Q}$ can be obtained through the arcsin law \cite{Jacovitti}
\begin{equation}
\mathbf{C}_{\mathbf{s}_\mathcal{Q}}=\frac{2}{\pi}\left(\text{sin}^{-1}(\mathbf{K}\mathbf{C}_{\mathbf{s}}^R\mathbf{K})+j\text{sin}^{-1}(\mathbf{K}\mathbf{C}_{\mathbf{s}}^I\mathbf{K})\right).
\label{equ_arcsin}
\end{equation}
The problem we are interested in solving in this work is to
cost-effectively estimate the channel parameters in $\mathbf{H}'$.

\section{Channel Estimation for Uplink 1-bit Oversampled MIMO}

In a standard uplink implementation, the channel state information
(CSI) is estimated at the BS and then used to detect the data
symbols transmitted from the $N_t$ users. Each transmission block is
divided into two sub-blocks: one for pilots and another for the data
symbols. Pilots are either located at the beginning of each block or
spread according to a desired pattern \cite{Coleri}. During the
training phase, each terminal simultaneously transmits $\tau$ pilot
symbols to the BS, which yields
\begin{equation}
\mathbf{y}_p=\mathbf{H}\mathbf{x}_p+\mathbf{n}_p.
\label{equ_pilot_model}
\end{equation}
Vectorizing (\ref{equ_pilot_model}) we get
\begin{equation}
\begin{aligned}
\mathbf{y}_p&=(\mathbf{x}_p^T\otimes \mathbf{I}_{N_r})\text{vec}(\mathbf{H})+\mathbf{n}_p\\&=[\mathbf{x}_p^T\otimes\mathbf{I}_{N_r}\otimes\mathbf{Z}(\mathbf{I}_{\tau}\otimes\mathbf{u})]\text{vec}(\mathbf{H}'\otimes\mathbf{I}_\tau)+\mathbf{n}_p\\&=\mathbf{\Phi}_p\mathbf{h'}+\mathbf{n}_p,
\end{aligned}
\label{equ_simplified_model}
\end{equation}
where $\mathbf{h'}=\text{vec}(\mathbf{H'})$ and the equivalent pilot matrix
\begin{equation}
\begin{aligned}
\mathbf{\Phi}_p &= [\mathbf{x}_p^T\otimes\mathbf{I}_{N_r}\otimes\mathbf{Z}(\mathbf{I}_\tau\otimes\mathbf{u})]\\&[\mathbf{I}_{N_t}\otimes(\mathbf{e}_1\otimes\mathbf{I}_{N_r}\otimes\mathbf{e}_1+\dots+\mathbf{e}_\tau\otimes\mathbf{I}_{N_r}\otimes\mathbf{e}_\tau)].
\end{aligned}
\end{equation}
The vector $\mathbf{x}_p\in\mathbb{C}^{\tau N_t\times1}$ contains
the transmitted pilots and $\mathbf{e}_n\in\mathbb{R}^{\tau\times1}$
represents a column vector with a one in the $n$th element and zeros
elsewhere. After processing by 1-bit ADCs, the quantized signal can
be expressed as
\begin{equation}
\mathbf{y}_{\mathcal{Q}_p}= \mathcal{Q}(\mathbf{\Phi}_p\mathbf{h'} + \mathbf{n}_p)=\tilde{\mathbf{\Phi}}_p\mathbf{h'} + \tilde{\mathbf{n}}_p,
\label{equ_linear}
\end{equation}
where $\tilde{\mathbf{\Phi}}_p = \mathbf{A}_p\mathbf{\Phi}_p\in\mathbb{C}^{M\tau N_r \times N_tN_r}$ and $\tilde{\mathbf{n}}_p = \mathbf{A}_p\mathbf{n}_p+\mathbf{n}_q\in\mathbb{C}^{M\tau N_r \times 1}$. The vector $\mathbf{n}_q$ is the statistically equivalent quantization noise\footnote{In this paper, we assume the quantization noise $\mathbf{n}_q$ is Gaussian distributed with zero mean and covariance $\mathbf{C}_{\mathbf{n}_q}$.} with covariance matrix $\mathbf{C}_{\mathbf{n}_q} = \mathbf{C}_{\mathbf{y}_{\mathcal{Q}_p}}-\mathbf{A}_p\mathbf{C}_{\mathbf{y}_p}\mathbf{A}_p^H$. The matrix $\mathbf{A}_p\in\mathbb{R}^{M\tau N_r \times M\tau N_r}$ is the Bussgang-based linear operator chosen independently from $\mathbf{y}_p$ and is given by
\begin{equation}
\mathbf{A}_p=\mathbf{C}_{\mathbf{y}_p\mathbf{y}_{\mathcal{Q}_p}}^H\mathbf{C}_{\mathbf{y}_p}^{-1}=\sqrt{\frac{2}{\pi}}\mathbf{K},
\label{equ_ap}
\end{equation}
where $\mathbf{C}_{\mathbf{y}_p\mathbf{y}_{\mathcal{Q}_p}}$ denotes the cross-correlation matrix between the received signal $\mathbf{y}_p$ and its quantized signal $\mathbf{y}_{\mathcal{Q}_p}$
\begin{equation}
\mathbf{C}_{\mathbf{y}_p\mathbf{y}_{\mathcal{Q}_p}}=\sqrt{\frac{2}{\pi}}\mathbf{K}\mathbf{C}_{\mathbf{y}_p}, \quad\text{with}\quad \mathbf{K} = \text{diag}(\mathbf{C}_{\mathbf{y}_p})^{-\frac{1}{2}}.
\label{equ_cap}
\end{equation}
The formulas of (\ref{equ_ap}) and (\ref{equ_cap}) involve the auto-correlation matrix $\mathbf{C}_{\mathbf{y}_p}$:
\begin{equation}
\mathbf{C}_{\mathbf{y}_p}=\mathbf{\Phi}_p\mathbf{R}_\mathbf{h'}\mathbf{\Phi}_p^H+\mathbf{C}_{\mathbf{n}_p},
\label{equ_Cy}
\end{equation}
where $\mathbf{R}_\mathbf{h'}=E\{\mathbf{h'}\mathbf{h'}^H\}$. We note that more sophisticated parameter estimation techniques can be adapted for use with 1-bit quantization \cite{smce,smtvb,jpais,armo,badstbc,baplnc,intadap,inttvt,ccmmwf,wlmwf,jio,jidf,sjidf,jidfecho,jiols,jiomimo,jiostap,jiostap,jiodoa,dce,locsme,okspme,l1stap,lrcc}.
\subsection{Noise covariance matrix $\mathbf{C}_{\MakeLowercase{\mathbf{n}_p}}$}

With (\ref{equ_noise}) the auto-correlation matrix
$\mathbf{C}_{\mathbf{n}_p}$ in (\ref{equ_Cy}) is calculated as
    \begin{equation}
    \mathbf{C}_{\mathbf{n}_p} = \sigma_n^2(\mathbf{I}_{N_r}\otimes \mathbf{GG}^H).
    \label{equ_C_n}
    \end{equation}
    For non-oversampled system ($M=1$), (\ref{equ_C_n}) is reduced to
    \begin{equation}
    \mathbf{C}_{\mathbf{n}_p} = \sigma_n^2\mathbf{I}_{\tau N_r}.
    \end{equation}
However, for oversampled system ($M\geq2$) (\ref{equ_C_n}) cannot be
further simplified due to the correlation of oversampled samples.
The off-diagonal elements will appear in the matrix of
$\mathbf{GG}^H$. 

\subsection{Standard LS Channel Estimator}
The work in \cite{Risi} has proposed the standard LS estimator for 1-bit non-oversampled systems. Similar to this, this estimator is extended to oversampled systems, which can be computed according to
\begin{equation}
\begin{aligned}
\hat{\mathbf{h'}}_{\text{Standard LS}}&=\argmin_{\bar{\mathbf{h'}}} ||\mathbf{y}_{\mathcal{Q}_p}-\mathbf{\Phi}_p\bar{\mathbf{h'}}||^2\\&=(\mathbf{\Phi}_p^H\mathbf{\Phi}_p)^{-1}\mathbf{\Phi}_p^H\mathbf{y}_{\mathcal{Q}_p}.
\end{aligned}
\end{equation}
The advantage of this estimator is that no a priori information is needed at the receiver. However, the issue with this estimator, when applied with 1-bit quantization, is that the channel estimate $\hat{\mathbf{h'}}$ scales with the amplitude associated with the quantizer, which then corresponds to a biased estimation.

\subsection{LRA-LS Channel Estimator}
Based on the Bussgang decomposition, the LS estimate is proposed for the linear equivalent system model in (\ref{equ_linear}). The LRA-LS channel estimator is obtained by solving the following optimization problem:
\begin{equation}
\begin{aligned}
\hat{\mathbf{h'}}_{\text{LRA-LS}}&=\argmin_{\bar{\mathbf{h'}}} ||\mathbf{y}_{\mathcal{Q}_p}-\tilde{\mathbf{\Phi}}_p\bar{\mathbf{h'}}||^2\\&=(\tilde{\mathbf{\Phi}}_p^H\tilde{\mathbf{\Phi}}_p)^{-1}\tilde{\mathbf{\Phi}}_p^H\mathbf{y}_{\mathcal{Q}_p}.
\end{aligned}
\label{equ_LS}
\end{equation}
Compared to the standard LS channel estimator, the proposed estimator has taken $\mathbf{R}_\mathbf{h'}$ into consideration in order to obtain the linear operator $\mathbf{A}_p$.

\subsection{LRA-LMMSE Channel Estimator}
The LMMSE channel estimator has the advantage of superior MSE performance to that of the LS channel estimator. Based on the statistically equivalent linear model in (\ref{equ_linear}), the oversampling based LRA-LMMSE channel estimator is proposed. The optimal filter is given by
\begin{equation}
\begin{aligned}
\mathbf{W}_{\text{LMMSE}} &=\argmin_{\mathbf{W}} E\{||\mathbf{h'}-\mathbf{W}\mathbf{y}_{\mathcal{Q}_p}||^2\}\\&= \mathbf{R}_\mathbf{h'}\tilde{\mathbf{\Phi}}^H\mathbf{C}_{\mathbf{y}_{\mathcal{Q}_p}}^{-1},
\end{aligned}
\end{equation}
where
\begin{equation}
\mathbf{C}_{\mathbf{y}_{\mathcal{Q}_p}} = \frac{2}{\pi}\left(\text{sin}^{-1}(\mathbf{K}\mathbf{C}_{\mathbf{y}_p}^R\mathbf{K})+j\text{sin}^{-1}(\mathbf{K}\mathbf{C}_{\mathbf{y}_p}^I\mathbf{K})\right).
\end{equation}
The resulting LRA-LMMSE channel estimator is then
\begin{equation}
\begin{aligned}
\hat{\mathbf{h'}}_{\text{LRA-LMMSE}} = \mathbf{R}_\mathbf{h'}\tilde{\mathbf{\Phi}}^H\mathbf{C}_{\mathbf{y}_{\mathcal{Q}_p}}^{-1}\mathbf{y}_{\mathcal{Q}_p}.
\end{aligned}
\label{equ_blmmse}
\end{equation}
\begin{IEEEproof}
    See Appendix A.
\end{IEEEproof}
Note that when $M=1$, (\ref{equ_blmmse}) reduces to the same as that of the BLMMSE channel estimator in \cite{Li}.

\subsection{LRA-LMS Channel Estimator}
LMS is the most widely used adaptive algorithm and has been adopted in various applications like system identification and channel equalization. In addition, LMS has robust performance and a low cost of implementation. Based on the linear equivalent model in (\ref{equ_linear}), an LRA-LMS channel estimator for 1-bit oversampled systems is devised.

\begin{figure}[!htbp]
    \centering
    \begin{tikzpicture}

	\draw [<->,thick] (0,0.7) node (yaxis) [above] {}
	|- (7.3,0) node (xaxis) [right] {$t$};
	\draw[step=0.5cm, line width=0.1mm, black] (0,0) grid (7,0.5);
	
	\draw [<->,thick] (0,-0.3) node (yaxis) [above] {}
	|- (7.3,-1) node (xaxis) [right] {$t$};
	\draw[step=0.5cm, line width=0.1mm, black] (0,-1) grid (7,-0.5);
	
	
	\node at (3.5,-1.25) {\tiny \textbullet};
	\node at (3.5,-1.45) {\tiny \textbullet};
	\node at (3.5,-1.65) {\tiny \textbullet};
	
	\draw [<->,thick] (0,-1.8) node (yaxis) [above] {}
	|- (7.3,-2.5) node (xaxis) [right] {$t$};
	\draw[step=0.5cm, line width=0.1mm, black] (0,-2.5) grid (7,-2);
	
	\draw [thick, decorate, decoration={brace, amplitude=10pt,mirror},xshift=0pt,yshift=-2pt]
	(4,-2.5) -- (7,-2.5) node [black,midway,yshift=-0.3cm,below] 
	{\footnotesize $Ml_{\textrm{win}}=6$};
	
	\fill [orange, opacity=0.2] (0,0) rectangle (1,0.5);
	\fill [orange, opacity=0.2] (1,0) rectangle (2,0.5);
	\fill [orange, opacity=0.2] (2,0) rectangle (3,0.5);

	\fill [orange, opacity=0.2] (1,-1) rectangle (2,-0.5);
	\fill [orange, opacity=0.2] (2,-1) rectangle (3,-0.5);
	\fill [orange, opacity=0.2] (3,-1) rectangle (4,-0.5);
	
	\fill [orange, opacity=0.2] (4,-2.5) rectangle (5,-2);
	\fill [orange, opacity=0.2] (5,-2.5) rectangle (6,-2);
	\fill [orange, opacity=0.2] (6,-2.5) rectangle (7,-2);
	
	\node at (-0.5,0.3) {1st};
	\node at (-0.5,-0.7) {2nd};
	\node at (-0.5,-2.2) {Last};
\end{tikzpicture}
    \caption{Illustration of the sliding window at each receive antenna when $l_{\text{win}}=3$ and $M=2$, where $l_{\text{win}}$ is the window length representing the number of symbols sampled at the Nyquist (symbol) rate.}
    \label{fig:sliding_window}
\end{figure}
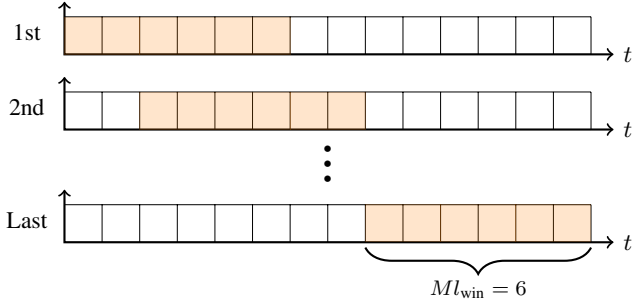

Since for large-scale MIMO with $N_r\gg N_t$, in order to reduce the computational complexity the multiplications and divisions involving large matrices, whose dimensions contain $N_r$ elements, need to be avoided. For this reason, we concentrate on the channel from $N_t$ users to only one receive antenna $n_r$ and the received quantized signal is modelled as
\begin{equation}
\mathbf{y}^{n_r}_{\mathcal{Q}_p} = \tilde{\mathbf{\Phi}}_p^{n_r}\mathbf{h'}^{n_r} + \tilde{\mathbf{n}}_p^{n_r},
\label{equ_sys_LMS}
\end{equation}
where $\mathbf{y}^{n_r}_{\mathcal{Q}_p}=[y^{n_r}_{\mathcal{Q}_p}(1),y^{n_r}_{\mathcal{Q}_p}(2),...,y^{n_r}_{\mathcal{Q}_p}(M\tau)]^T$ and $\mathbf{h'}^{n_r}\in \mathbb{C}^{N_t\times 1}$ is the $n_r$th row of $\mathbf{H'}$. Different from $\tilde{\mathbf{\Phi}}_p$ in (\ref{equ_linear}), $\tilde{\mathbf{\Phi}}_p^{n_r}\in \mathbb{C}^{M\tau\times N_t}$ is an equivalent pilot matrix to the $n_r$th receive antenna. The sliding window based technique \cite{Shao3} (shown in Fig. \ref{fig:sliding_window}) is applied, which combines the adjacent symbol-rate-sampled symbols together to estimate the instantaneous channel parameters, since in oversampled systems the interference from adjacent symbol-rate-sampled symbols should be considered. The first window contains the first $Ml_{\text{win}}$ oversampled samples and the second contains the next $Ml_{\text{win}}$ samples until the last window. Note that only one symbol-rate-sampled symbol (or $M$ oversampled samples) is shifted for the subsequent window.

Based on (\ref{equ_sys_LMS}), the received signal at the $n$th window can be expressed as
\begin{equation}
\mathbf{y}^{n_r}_{\mathcal{Q}_p}(n) = \tilde{\mathbf{\Phi}}_p^{n_r}(n)\mathbf{h'}^{n_r} + \tilde{\mathbf{n}}_p^{n_r}(n),
\end{equation}
where $\mathbf{y}^{n_r}_{\mathcal{Q}_p}(n)=[y^{n_r}_{\mathcal{Q}_p}(M(n-1)+1),\dots,y^{n_r}_{\mathcal{Q}_p}(M(n-1)+Ml_{\text{win}})]^T$ and $\tilde{\mathbf{\Phi}}^{n_r}_p(n)=\mathbf{A}^{n_r}_p(n)\mathbf{\Phi}^{n_r}_p(n)\in\mathbb{C}^{Ml_{\text{win}} \times N_t}$ contains the transmit pilot sequences in the $n$th window.

The optimization problem that leads to the proposed LRA-LMS channel estimation algorithm can be stated as
\begin{equation}
\resizebox{\columnwidth}{!}{$\displaystyle
\hat{\mathbf{h'}}^{n_r}_{\text{LRA-LMS}}(n)=\argmin_{\bar{\mathbf{h'}}^{n_r}(n)}\sum_{n=1}^{\tau-l_{\text{win}}+1}||\mathbf{y}^{n_r}_{\mathcal{Q}_p}(n)-\tilde{\mathbf{\Phi}}^{n_r}_p(n)\bar{\mathbf{h'}}^{n_r}(n)||^2,$}
\label{equ:objective}
\end{equation}
where $\bar{\mathbf{h'}}^{n_r}(n)$ is the instantaneous estimate of $\mathbf{h'}^{n_r}$ in the $n$th window.

Taking the partial derivative of the objective function in (\ref{equ:objective}) with respect to $\bar{\mathbf{h'}}^{n_r}(n)^H$, we obtain
\begin{equation}
\begin{aligned}
&\qquad\frac{\partial \sum_{n=1}^{\tau-l_{\text{win}}+1}||\mathbf{y}^{n_r}_{\mathcal{Q}_p}(n)-\tilde{\mathbf{\Phi}}^{n_r}_p(n)\bar{\mathbf{h'}}^{n_r}(n)||^2}{\partial \bar{\mathbf{h'}}^{n_r}(n)^H} \\&=\sum_{n=1}^{\tau-l_{\text{win}}+1}-\tilde{\mathbf{\Phi}}^{n_r}_p(n)^H(\mathbf{y}^{n_r}_{\mathcal{Q}_p}(n)-\tilde{\mathbf{\Phi}}^{n_r}_p(n)\bar{\mathbf{h'}}^{n_r}(n))\\&=\sum_{n=1}^{\tau-l_{\text{win}}+1}-\tilde{\mathbf{\Phi}}^{n_r}_p(n)^H\mathbf{e}^{n_r}(n).
\end{aligned}
\label{equ:derivation}
\end{equation}
The recursion of the proposed LRA-LMS algorithm is
\begin{equation}
\begin{aligned}
\bar{\mathbf{h'}}^{n_r}(n+1) = \bar{\mathbf{h'}}^{n_r}(n) + \mu& \tilde{\mathbf{\Phi}}^{n_r}_p(n)^H\mathbf{e}^{n_r}(n),\\&\hspace{0.7em}n=1,\ldots,\tau-l_{\text{win}}+1,
\end{aligned}
\label{equ_recursion}
\end{equation}
where the constant step size $\mu$ fulfills
\begin{equation}
0<\mu<\frac{2}{\gamma_{\max}}.
\label{equ_condition}
\end{equation}
$\gamma_{\max}$ is the largest eigenvalue of $\mathbf{C}_{\tilde{\mathbf{\Phi}}^{n_r}_p(n)}$, which is $E\{\tilde{\mathbf{\Phi}}^{n_r}_p(n)\tilde{\mathbf{\Phi}}^{n_r}_p(n)^H\}$.
\begin{IEEEproof}
    See Appendix B.
\end{IEEEproof}

\begin{figure}[!htbp]
    \centering
    \pgfplotsset{every axis label/.append style={font=\footnotesize},
	every tick label/.append style={font=\footnotesize},
}

\definecolor{mycolor1}{rgb}{1.00000,0.00000,1.00000}%
\definecolor{mycolor2}{rgb}{0.50000,0.00000,0.50000}%

\begin{tikzpicture}

\begin{axis}[%
width=.8\columnwidth,
height=.5\columnwidth,
at={(0.758in,0.603in)},
scale only axis,
xmin=10,
xmax=80,
xlabel style={font=\footnotesize},
xlabel={Pilot length $\tau$},
xtick=data,
xmajorgrids,
ymin= -10,
ymax=-3,
yminorticks=true,
ylabel style={font=\footnotesize},
ylabel={Normalized MSE (dB)},
ymajorgrids,
yminorgrids,
axis background/.style={fill=white},
legend entries={$1$ iteration,
	$2$ iterations,
	$3$ iterations},
legend style={at={(0.6,0.67)},anchor=south west,legend cell align=left,align=left,draw=white!15!black,font=\scriptsize}
]

\addplot [color=red,solid,line width=1.0pt,mark=square,mark options={solid},mark size=4pt]
table[row sep=crcr]{%
	10	-3.37685297697237\\
	20	-5.83619644865242\\
	30	-6.77694038324465\\
	40	-7.08067046876470\\
	50	-7.15910246688600\\
	60	-7.15283327974977\\
	70	-7.14161571157804\\
	80  -7.16222880566842\\
};
\addlegendentry{M=1, $\mu=0.3$};

\addplot [color=blue,solid,line width=1.0pt,mark=square,mark options={solid},mark size=4pt]
table[row sep=crcr]{%
	10	-3.83810476755854\\
	20	-6.66352913400616\\
	30	-7.87590861022607\\
	40	-8.29951042754197\\
	50  -8.41696451424772\\
	60  -8.47617701708517\\
	70	-8.48617749799269\\
	80  -8.49409028437698\\
};
\addlegendentry{M=2, $\mu=0.18$};

\addplot [color=green,solid,line width=1.0pt,mark=square,mark options={solid},mark size=4pt]
table[row sep=crcr]{%
	10	-3.79489510124494\\
	20	-6.57237845766499\\
	30	-7.82352936895983\\
	40	-8.28878094814380\\
	50  -8.46900676261289\\
	60	-8.51187274717017\\
	70  -8.57749470651889\\
	80  -8.55837554123846\\
};
\addlegendentry{M=3, $\mu=0.12$};

\end{axis}

\end{tikzpicture}%
    \caption{Convergence of the LRA-LMS channel estimator with $N_t = 8$ and $N_r = 64$ at SNR = 20dB.}
    \label{fig:convergence}
\end{figure}
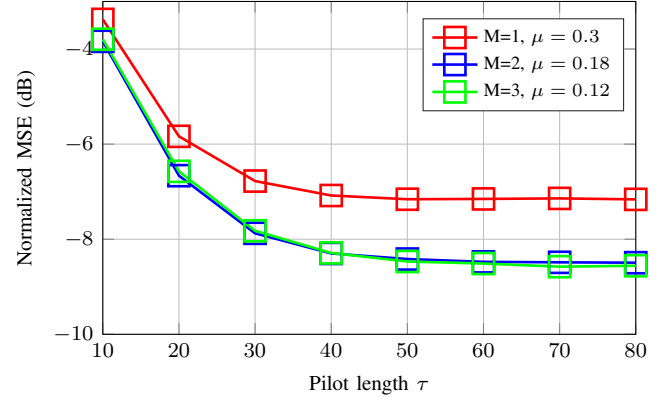

The proposed adaptive channel estimator is summarized in Algorithm \ref{alg:receiver}, where $\mathbf{x}_p(n)\in\mathbb{C}^{l_{\text{win}}N_t \times 1}$ contains the pilot symbols in the $n$th window. Both $\mathbf{e}'_n\in\mathbb{R}^{l_{\text{win}}\times 1}$ and $\mathbf{e}''_n\in\mathbb{R}^{N_r\times 1}$ represent all-zero column vectors except that the $n$th elements are ones.
\begin{algorithm}
    \caption{Proposed LRA-LMS Channel Estimator}
    \begin{algorithmic}[1]    
        \Parameters{$\mu$: forgetting factor}
        \vspace{1mm}
        \Initialization{$\mathbf{h'}^{n_r}(1)=\mathbf{0}_{N_t\times 1}$}
        \vspace{1mm}
        \Iteration{}
        \vspace{-4mm}
        \For{$n_r=1:N_r$}
        \For{$n=1:\tau-l_{\text{win}}+1$}
        \State $\begin{multlined}[]
        \mathbf{\Phi}^{n_r}_p(n) = [\mathbf{x}_p^T(n)\otimes\mathbf{Z}(\mathbf{I}_{l_{\text{win}}}\otimes\mathbf{u})]\\\hspace{1.2cm}[\mathbf{I}_{N_t}\otimes(\mathbf{e}'_1\otimes\mathbf{e}'_1+\dots+\mathbf{e}'_{l_{\text{win}}}\otimes\mathbf{e}'_{l_{\text{win}}})];
        \end{multlined}$
        \State $\mathbf{C}^{n_r}_{\mathbf{y}_p}(n)=\mathbf{\Phi}^{n_r}_p(n)\mathbf{\Phi}^{n_r}_p(n)^H+\sigma_n^2\mathbf{GG}^H$;
        \State $\mathbf{A}^{n_r}_p(n)=\sqrt{\frac{2}{\pi}}\text{diag}(\mathbf{C}^{n_r}_{\mathbf{y}_p}(n))^{-\frac{1}{2}}$;
        \State $\tilde{\mathbf{\Phi}}^{n_r}_p(n)=\mathbf{A}^{n_r}_p(n)\mathbf{\Phi}^{n_r}_p(n)$;
        \State $\mathbf{e}^{n_r}(n) = \mathbf{y}^{n_r}_{\mathcal{Q}_p}(n)-\tilde{\mathbf{\Phi}}^{n_r}_p(n)\mathbf{h'}^{n_r}(n)$;
        \vspace{1mm}
        \State $\mathbf{h'}^{n_r}(n+1) = \mathbf{h'}^{n_r}(n) + \mu \tilde{\mathbf{\Phi}}^{n_r}_p(n)^H\mathbf{e}^{n_r}(n)$;
        \vspace{1mm}
        \EndFor
        \EndFor
    \end{algorithmic}
    \label{alg:receiver}
\end{algorithm}
Fig. \ref{fig:convergence} shows the convergence performance of the proposed LRA-LMS channel estimator for each receive antenna. The proposed estimator achieves its steady state after $\tau=40$.

\subsection{Complexity Analysis}
\begin{figure}[!htbp]
    \centering
    \pgfplotsset{every axis label/.append style={font=\footnotesize},
	every tick label/.append style={font=\footnotesize},
}

\definecolor{mycolor1}{rgb}{1.00000,0.00000,1.00000}%

\begin{tikzpicture}
[spy using outlines={black,rectangle,magnification=3,size=1.5cm, connect spies}]

\begin{axis}[%
width=.8\columnwidth,
height=.5\columnwidth,
at={(0.758in,0.603in)},
scale only axis,
xmin=20,
xmax=100,
xlabel style={font=\footnotesize},
xlabel={Number of receive antennas $N_r$},
xtick=data,
xmajorgrids,
ymode=log,
ymin=1e6,
ymax=1e13,
yminorticks=true,
ylabel style={font=\footnotesize},
ylabel={Total number of complex operations},
ymajorgrids,
yminorgrids,
axis background/.style={fill=white},
legend style={at={(0.98,0.68)},anchor=north east,legend cell align=left,align=left,draw=white!15!black,font=\scriptsize}
]

\addplot [color=green,solid,line width=1.0pt,mark=o,mark options={solid},mark size=4pt]
table[row sep=crcr]{%
	100	3090655456000.00\\
	90	2253271833600.00\\
	80  1582706374400.00\\
	70	1060428774400.00\\
	60	667908729600.000\\
	50	386615936000.000\\
	40	198020089600.000\\
	30  83590886400.0000\\
	20  24798022400.0000\\
};
\addlegendentry{Standard LS};

\addplot [color=blue,solid,line width=1.0pt,mark=diamond,mark options={solid},mark size=4pt]
  table[row sep=crcr]{%
  	100	3205923960400.00\\
  	90	2337308368800.00\\
  	80  1641733084400.00\\
  	70	1099976603200.00\\
  	60	692817421200.000\\
  	50	401034034400.000\\
  	40	205404938800.000\\
  	30  86708630400.0000\\
  	20  25723605200.0000\\
};
\addlegendentry{LRA-LS};

\addplot [color=red,solid,line width=1.0pt,mark=triangle,mark options={solid},mark size=4pt]
table[row sep=crcr]{%
	100	3679328599600.00\\
	90	2682435630480.00\\
	80  1884140405360.00\\
	70	1262382124240.00\\
	60	795099987120.000\\
	50	460233194000.000\\
	40	235720944880.000\\		
	30  99502439760.0000\\		
	20  29516878640.0000\\										
};
\addlegendentry{LRA-LMMSE};

\addplot [color=mycolor1,solid,line width=1.0pt,mark=square,mark options={solid},mark size=4pt]
table[row sep=crcr]{%
	100	34552800\\
	90	31097520\\
	80  27642240\\
	70	24186960\\
	60	20731680\\
	50	17276400\\
	40	13821120\\	
	30  10365840\\
	20  6910560\\
};
\addlegendentry{LRA-LMS};

\end{axis}

\spy on (2.77,4.55) in node [left] at (5,3.6);

\end{tikzpicture}%
    \caption{Computational complexity comparison between different channel estimators in an oversampled system $M=3$ with $\tau = 20$, $l_{\text{win}}=3$ and $N_t = 8$.}
    \label{fig:Complexity}
\end{figure}
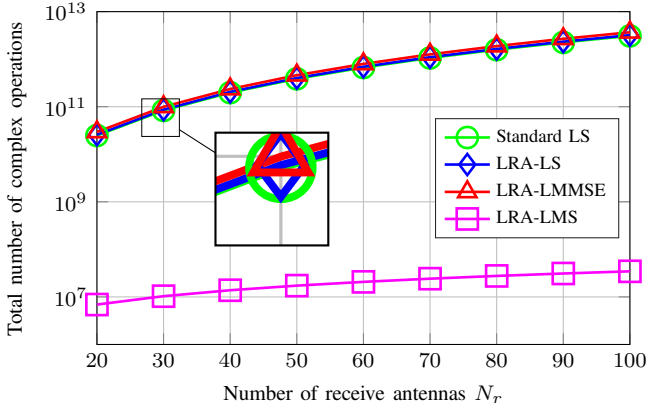

The computational complexities of the proposed channel estimators are compared in this subsection. For the sake of simplification and a fair comparison among the estimators, we assume $\mathbf{R}_\mathbf{h'}$ is an identity matrix. Table \ref{tab:complexity} shows the total required complex additions/subtractions and multiplications/divisions for obtaining the channel estimate $\hat{\mathbf{h'}}$. More intuitively, Fig. \ref{fig:Complexity} shows the total number of complex operations, which is a sum of complex additions and multiplications, as a function of the number of receive antennas $N_r$. Compared to other channel estimators, the LRA-LMS channel estimator consumes the lowest computational cost since there are no matrix inversions or large matrix multiplications in the algorithm. The comparisons in terms of MSE performance are shown in the simulations section.

\begin{table*}[!htbp]
    \caption{Computational complexity of different channel estimators}
    \label{tab:complexity}
    \centering
    \begin{tabular}{c||c|c}
        \hline
        & Complex Additions/Subtractions & Complex Multiplications/Divisions\\ \hline\hline
        Standard LS      &   \begin{tabular}{@{}c@{}c@{}}$N_r^3N_t^2(N_t+M\tau^3+2M\tau)$ \\ $-N_r^2N_t(M\tau+N_t)-2N_rN_t$\\
            $+M\tau^2(M\tau-1)$\end{tabular}   &
        \begin{tabular}{@{}c@{}c@{}}$N_r^3N_t^2(N_t+2M\tau+\tau^3M)$ \\ $+N_r^2\tau[\tau N_t^2+(\tau^2+1)MN_t+\tau^2+(1+M)\tau]$\\
            $+2N_rN_t+M\tau^2(1+\tau)$\end{tabular}\\ \hline
        LRA-LS      &   \begin{tabular}{@{}c@{}c@{}}$N_r^3N_t[N_t^2+(M\tau^3+2M\tau)N_t+2M^2\tau^2]$ \\ $-N_r^2N_t(2M\tau+N_t)-2N_rN_t$\\
            $+M\tau^2(M\tau-M-1+3M^2\tau)$\end{tabular}   &
        \begin{tabular}{@{}c@{}c@{}}$N_r^3N_t[N_t^2+(M\tau^3+2M\tau)N_t+2M^2\tau^2]$ \\ $+N_r^2\tau[\tau N_t^2+(\tau^2+1)MN_t+\tau^2+(1+M+2M^2)\tau]$\\
            $+2N_r(M\tau+N_t)+3M^3\tau^3+M\tau^2(1+\tau)$\end{tabular}\\ \hline
        LRA-LMMSE       & \begin{tabular}{@{}c@{}c@{}}$N_r^3[M\tau^3N_t^2+3M^2\tau^2N_t+M^3\tau^3]$ \\ $-2N_r^2M\tau N_t-N_r(M\tau+N_t)$\\
            $+M\tau^2(M\tau-1+3M^2\tau-M)$\end{tabular}     & \begin{tabular}{@{}c@{}c@{}}$N_r^3[M\tau^3N_t^2+3M^2\tau^2N_t+M^3\tau^3]$ \\ $+N_r^2\tau[\tau N_t^2+(\tau^2+1)MN_t+\tau^2+(1+\tau+M+5M^2)\tau]$\\
            $+3M\tau N_r+3M^3\tau^3+M\tau^2(1+\tau)$\end{tabular} \\ \hline
        LRA-LMS & \begin{tabular}{@{}c@{}}$N_r(\tau-l_{\text{win}}+1)[l_{\text{win}}^2M(2MN_t-M-1)$ \\ $+l_{\text{win}}^3M(3M^2+M+N_t^2)]$\end{tabular} & \begin{tabular}{@{}c@{}c@{}}$N_r(\tau-l_{\text{win}}+1)[N_t+2l_{\text{win}}M(1+N_t)$ \\ $+l_{\text{win}}^2(1+N_t^2+2M+2M^2N_t+2M^2)$\\
            $+l_{\text{win}}^3(N_t^2M+1+MN_t+M+3M^3)]$\end{tabular}  \\ \hline
    \end{tabular}
\end{table*}

\subsection{Estimation of $\mathbf{R}_\MakeLowercase{\mathbf{h'}}$}
In practical environments, there is no prior information about $\mathbf{R}_\mathbf{h'}$ at the receiver. In this subsection, an adaptive technique is proposed to recursively estimate $\mathbf{R}_\mathbf{h'}$ as
\begin{equation}
\hat{\mathbf{R}}_\mathbf{h'}(n+1) = \lambda\hat{\mathbf{R}}_\mathbf{h'}(n) + (1-\lambda) \hat{\mathbf{h'}}(n)\hat{\mathbf{h'}}(n)^H,\hspace{0.3cm}n=1,\ldots,\tau,
\label{equ_Rh}
\end{equation}
where $\lambda$ is the forgetting factor and $\hat{\mathbf{h'}}(n)$ is the channel estimate at the Nyquist time instant n. Consider the system model
\begin{equation}
\begin{aligned}
\mathbf{y}_\mathcal{Q}(n)&=\mathcal{Q}(\mathbf{H}\mathbf{x}(n)+\mathbf{n}(n))\\&=\mathcal{Q}((\mathbf{x'}_p^T(n)\otimes\mathbf{I}_{N_r}\otimes\mathbf{Z'u})\mathbf{h'}+\mathbf{n}(n)),
\end{aligned}
\label{equ_sysmodeln}
\end{equation}
where $\mathbf{y}_\mathcal{Q}(n)$ and $\mathbf{n}(n)$ are column vectors with size $MN_r\times1$. Different from $\mathbf{x}_p(n)$ in Algorithm \ref{alg:receiver}, $\mathbf{x'}_p(n)\in\mathbb{C}^{N_t\times1}$ contains pilot symbols from $N_t$ terminals at time instant n. $\mathbf{Z'}\in\mathbb{R}^{M \times M}$ is a simplified version of $\mathbf{Z}$ with $N=1$. The instantaneous estimate of $\mathbf{h'}$ is calculated as
\begin{equation}
\hat{\mathbf{h'}}(n) =  (\mathbf{x'}_p^T(n)\otimes\mathbf{I}_{N_r}\otimes\mathbf{Z'u})^+\mathbf{y}_\mathcal{Q}(n),
\label{equ_esth}
\end{equation}
where the initial guess of $\hat{\mathbf{R}}_\mathbf{h'}(1)$ is an identity matrix by assuming channel parameters are uncorrelated and each has unit power.

\section{Cram\'er-Rao Bounds}
Unlike the works in \cite{Schluter,Stein}, which have proposed the CRBs for the unbiased estimators, the existing CRBs are extended suitable for the biased estimators. Two different types of CRBs are proposed depending on whether the prior information $\mathbf{R}_\mathbf{h'}$ is known at the receiver, namely Bayesian CRB with known $\mathbf{R}_\mathbf{h'}$ and general CRB with estimated $\mathbf{R}_\mathbf{h'}$.

\subsection{Bayesian Cram\'er-Rao Bounds}
Bayesian bounds on the fundamental limits of estimation are derived for non-oversampled and oversampled systems. Without loss of generality, we extend (\ref{equ_simplified_model}) considering the whole system and not just the pilots, and rewrite the complex-valued model in the following real-valued form
\begin{equation}
\begin{bmatrix}
\mathbf{y}^R \\
\mathbf{y}^I
\end{bmatrix}=\begin{bmatrix}
\mathbf{\Phi}^R &-\mathbf{\Phi}^I\\
\mathbf{\Phi}^I & \mathbf{\Phi}^R
\end{bmatrix}\begin{bmatrix}
\mathbf{h'}^R \\
\mathbf{h'}^I
\end{bmatrix}+\begin{bmatrix}
\mathbf{n}^R \\
\mathbf{n}^I
\end{bmatrix}.
\label{equ:system_model_real}
\end{equation}

Let $\tilde{\mathbf{h'}}=[\mathbf{h'}^R;\mathbf{h'}^I]$ be the unknown parameter vector, since the real and imaginary parts are independent, the Bayesian information matrix (BIM) \cite{Trees} for the quantized signal is defined as
\begin{equation}
\mathbf{J}_{\mathbf{y}_\mathcal{Q}}(\tilde{\mathbf{h'}})=\mathbf{J}_{\mathbf{y}_\mathcal{Q}^R}(\tilde{\mathbf{h'}})+\mathbf{J}_{\mathbf{y}_\mathcal{Q}^I}(\tilde{\mathbf{h'}}),
\label{equ_total_baysian}
\end{equation}
where
\begin{equation}
\resizebox{\columnwidth}{!}{$\displaystyle
[\mathbf{J}_{\mathbf{y}_\mathcal{Q}^{R/I}}(\tilde{\mathbf{h'}})]_{ij}\triangleq E_{\mathbf{y}_\mathcal{Q}^{R/I},\tilde{\mathbf{h'}}}\left\{\frac{\partial \ln p(\mathbf{y}_\mathcal{Q}^{R/I},\tilde{\mathbf{h'}})}{\partial [\tilde{\mathbf{h'}}]_i}\frac{\partial \ln p(\mathbf{y}_\mathcal{Q}^{R/I},\tilde{\mathbf{h'}})}{\partial [\tilde{\mathbf{h'}}]_j}\right\}$}
\label{equ_bayesian}
\end{equation}
with $[\tilde{\mathbf{h'}}]_i$ and $[\tilde{\mathbf{h'}}]_j$ being the elements of $\tilde{\mathbf{h'}}$. The expression in (\ref{equ_bayesian}) can be divided into two parts:
\begin{equation}
[\mathbf{J}_{\mathbf{y}_\mathcal{Q}^{R/I}}(\tilde{\mathbf{h'}})]_{ij} = [\mathbf{J}^D_{\mathbf{y}_\mathcal{Q}^{R/I}}(\tilde{\mathbf{h'}})]_{ij}+[\mathbf{J}^P_{\mathbf{y}_\mathcal{Q}^{R/I}}(\tilde{\mathbf{h'}})]_{ij},
\end{equation}
where
\begin{equation}
\resizebox{\columnwidth}{!}{$\displaystyle
[\mathbf{J}^D_{\mathbf{y}_\mathcal{Q}^{R/I}}(\tilde{\mathbf{h'}})]_{ij}\triangleq E_{\mathbf{y}_\mathcal{Q}^{R/I}\mid\tilde{\mathbf{h'}}}\left\{\frac{\partial \ln p(\mathbf{y}_\mathcal{Q}^{R/I}\mid\tilde{\mathbf{h'}})}{\partial [\tilde{\mathbf{h'}}]_i}\frac{\partial \ln p(\mathbf{y}_\mathcal{Q}^{R/I}\mid\tilde{\mathbf{h'}})}{\partial [\tilde{\mathbf{h'}}]_j}\right\}$}
\label{equ_Jd}
\end{equation}
\begin{equation}
[\mathbf{J}^P_{\mathbf{y}_\mathcal{Q}^{R/I}}(\tilde{\mathbf{h'}})]_{ij} \triangleq E_{\tilde{\mathbf{h'}}}\left\{\frac{\partial \ln p(\tilde{\mathbf{h'}})}{\partial [\tilde{\mathbf{h'}}]_i}\frac{\partial \ln p(\tilde{\mathbf{h'}})}{\partial [\tilde{\mathbf{h'}}]_j}\right\}.
\label{equ_h}
\end{equation}

To transform the real-valued $\mathbf{J}_{\mathbf{y}_\mathcal{Q}}(\tilde{\mathbf{h'}})$ back to the complex domain $\mathbf{J}_{\mathbf{y}_\mathcal{Q}}(\mathbf{h'})$, $\mathbf{J}_{\mathbf{y}_\mathcal{Q}}(\tilde{\mathbf{h'}})$ is defined with the following structure:
\begin{equation}
\mathbf{J}_{\mathbf{y}_\mathcal{Q}}(\tilde{\mathbf{h'}})=\begin{bmatrix}
\mathbf{J}^{RR}_{\mathbf{y}_\mathcal{Q}}(\tilde{\mathbf{h'}}) & \mathbf{J}^{RI}_{\mathbf{y}_\mathcal{Q}}(\tilde{\mathbf{h'}}) \\ \mathbf{J}^{IR}_{\mathbf{y}_\mathcal{Q}}(\tilde{\mathbf{h'}}) & \mathbf{J}^{II}_{\mathbf{y}_\mathcal{Q}}(\tilde{\mathbf{h'}})
\end{bmatrix}
\label{equ_Jmatrix}
\end{equation}
and apply the chain rule to get:
\begin{equation}
\mathbf{J}_{\mathbf{y}_\mathcal{Q}}(\mathbf{h'})=\frac{1}{4}(\mathbf{J}_{\mathbf{y}_\mathcal{Q}}^{RR}(\tilde{\mathbf{h'}})+\mathbf{J}_{\mathbf{y}_\mathcal{Q}}^{II}(\tilde{\mathbf{h'}}))+\frac{j}{4}(\mathbf{J}_{\mathbf{y}_\mathcal{Q}}^{RI}(\tilde{\mathbf{h'}})-\mathbf{J}_{\mathbf{y}_\mathcal{Q}}^{IR}(\tilde{\mathbf{h'}})),
\end{equation}
where $\mathbf{J}^{RR}_{\mathbf{y}_\mathcal{Q}}(\tilde{\mathbf{h'}})$, $\mathbf{J}^{RI}_{\mathbf{y}_\mathcal{Q}}(\tilde{\mathbf{h'}})$, $\mathbf{J}^{IR}_{\mathbf{y}_\mathcal{Q}}(\tilde{\mathbf{h'}})$ and $\mathbf{J}^{II}_{\mathbf{y}_\mathcal{Q}}(\tilde{\mathbf{h'}})$ have the same dimensions $N_rN_t\times N_rN_t$. The variance of the estimator $\hat{\mathbf{h'}}(\mathbf{y}_\mathcal{Q})$ is lower bounded by
\begin{equation}
Var\{\hat{h'}_i(\mathbf{y}_\mathcal{Q})\}\geq[\mathbf{J}_{\mathbf{y}_\mathcal{Q}}^{-1}(\mathbf{h'})]_{ii}.
\label{equ_bounds}
\end{equation}

\subsubsection{BIM for Non-oversampled Systems}
For non-oversampled systems, i.e, $M=1$, the covariance matrix of the equivalent noise vector $\mathbf{n}$ is $\mathbf{C}_\mathbf{n}=\sigma_n^2\mathbf{I}_{NN_r}$. With the independence of the real and imaginary parts, the log-likelihood function can be expressed as
\begin{equation}
\ln p(\mathbf{y}_\mathcal{Q}\mid\tilde{\mathbf{h'}})=\sum_{k=1}^{NN_r}[\ln p([\mathbf{y}_\mathcal{Q}^R]_k\mid\tilde{\mathbf{h'}})+\ln p([\mathbf{y}_\mathcal{Q}^I]_k\mid\tilde{\mathbf{h'}})]
\label{equ_LLR}
\end{equation}
with
\begin{equation}
p([\mathbf{y}_\mathcal{Q}^{R}]_k=\pm\frac{1}{\sqrt{2}}\mid\tilde{\mathbf{h'}})=Q\left(\mp\frac{[\mathbf{\Phi}^R\mathbf{h'}^R-\mathbf{\Phi}^I\mathbf{h'}^I]_k}{\sigma_n/\sqrt{2}}\right)
\end{equation}
\begin{equation}
p([\mathbf{y}_\mathcal{Q}^{I}]_k=\pm\frac{1}{\sqrt{2}}\mid\tilde{\mathbf{h'}})=Q\left(\mp\frac{[\mathbf{\Phi}^I\mathbf{h'}^R+\mathbf{\Phi}^R\mathbf{h'}^I]_k}{\sigma_n/\sqrt{2}}\right)
\end{equation}
where $Q(x) = \frac{1}{\sqrt{2\pi}}\int_x^\infty \exp(-\frac{u^2}{2})du$. Inserting (\ref{equ_LLR}) into (\ref{equ_Jd}), we obtain
\begin{equation}
\begin{aligned}
[\mathbf{J}^D_{\mathbf{y}_\mathcal{Q}}(\tilde{\mathbf{h'}})]_{ij} &= -E\left\{\frac{\partial^2\ln p(\mathbf{y}_\mathcal{Q}\mid\tilde{\mathbf{h'}})}{\partial [\tilde{\mathbf{h'}}]_i\partial [\tilde{\mathbf{h'}}]_j}\right\}\\&=[\mathbf{J}^D_{\mathbf{y}^R_\mathcal{Q}}(\tilde{\mathbf{h'}})]_{ij}+[\mathbf{J}^D_{\mathbf{y}^I_\mathcal{Q}}(\tilde{\mathbf{h'}})]_{ij}.
\end{aligned}
\label{equ_Jdd}
\end{equation}
With the derivative of the $Q(x)$ function, the real part in (\ref{equ_Jd}) $[\mathbf{J}^D_{\mathbf{y}^R_\mathcal{Q}}(\tilde{\mathbf{h'}})]_{ij}$is given by
\begin{equation}
\resizebox{\columnwidth}{!}{$\displaystyle\begin{aligned}
[\mathbf{J}^D_{\mathbf{y}^R_\mathcal{Q}}(\tilde{\mathbf{h'}})]_{ij} &= \sum_{k=1}^{NN_r}-E\left\{\frac{\partial^2\ln p([\mathbf{y}_\mathcal{Q}^R]_k\mid\tilde{\mathbf{h'}})}{\partial [\tilde{\mathbf{h'}}]_i\partial [\tilde{\mathbf{h'}}]_j}\right\}=\frac{1}{\pi\sigma_n^2}\\&\hspace{-2cm}\times\sum_{k=1}^{NN_r}\frac{\exp(-\frac{[\mathbf{\Phi}^R\mathbf{h'}^R-\mathbf{\Phi}^I\mathbf{h'}^I]_k^2}{\sigma_n^2/2})\frac{\partial[\mathbf{\Phi}^R\mathbf{h'}^R-\mathbf{\Phi}^I\mathbf{h'}^I]_k}{\partial [\tilde{\mathbf{h'}}]_i}\frac{\partial[\mathbf{\Phi}^R\mathbf{h'}^R-\mathbf{\Phi}^I\mathbf{h'}^I]_k}{\partial [\tilde{\mathbf{h'}}]_j}}{Q\left(\frac{[\mathbf{\Phi}^R\mathbf{h'}^R-\mathbf{\Phi}^I\mathbf{h'}^I]_k}{\sigma_n/\sqrt{2}}\right)Q\left(-\frac{[\mathbf{\Phi}^R\mathbf{h'}^R-\mathbf{\Phi}^I\mathbf{h'}^I]_k}{\sigma_n/\sqrt{2}}\right)}.
\end{aligned}$}
\label{equ_JDR}
\end{equation}
The derivation for the imaginary part $[\mathbf{J}^D_{\mathbf{y}^I_\mathcal{Q}}(\tilde{\mathbf{h'}})]_{ij}$ is analogous.

By assuming that $\tilde{\mathbf{h'}}$ is Gaussian distributed with zero mean and covariance matrix $\mathbf{C}_{\tilde{\mathbf{h'}}}=\frac{1}{2}\mathbf{I}_2\otimes\mathbf{C}_\mathbf{h'}$, $\ln p(\tilde{\mathbf{h'}})$ yields
\begin{equation}
\ln p(\tilde{\mathbf{h'}}) = -\frac{1}{2}N_rN_t\ln[(2\pi)^{2N_rN_t}\det(\mathbf{C}_{\tilde{\mathbf{h'}}})]-\frac{1}{2}\tilde{\mathbf{h'}}^T\mathbf{C}_{\tilde{\mathbf{h'}}}^{-1}\tilde{\mathbf{h'}}.
\label{equ_lnph}
\end{equation}
Substituting (\ref{equ_lnph}) into (\ref{equ_h}), we obtain
\begin{equation}
\mathbf{J}^P_{\mathbf{y}_\mathcal{Q}}(\tilde{\mathbf{h'}})=2\mathbf{J}^P_{\mathbf{y}^{R/I}_\mathcal{Q}}(\tilde{\mathbf{h'}})=2\mathbf{C}_{\tilde{\mathbf{h'}}}^{-1}.
\label{equ_Jp}
\end{equation}

Finally, the resulting BIM is the summation of (\ref{equ_Jdd}) and (\ref{equ_Jp}) as described by
\begin{equation} \mathbf{J}_{\mathbf{y}_\mathcal{Q}}(\tilde{\mathbf{h'}})=\mathbf{J}^D_{\mathbf{y}_\mathcal{Q}}(\tilde{\mathbf{h'}})+\mathbf{J}^P_{\mathbf{y}_\mathcal{Q}}(\tilde{\mathbf{h'}}).
\end{equation}

\subsubsection{BIM for Oversampled Systems}
When $M\geq2$ the equivalent noise vector $\mathbf{n}$ consists of colored Gaussian noise samples. Computing $p(\mathbf{y}_\mathcal{Q}^{R/I}\mid\tilde{\mathbf{h'}})$ requires the orthant probabilities, which are not available or too difficult to compute. The authors in \cite{Stein,Stein2} have introduced a lower bounding technique on the Fisher information for real-valued system. To employ this lower bounding technique in the complex-valued system, the work of \cite{Schluter} has come out. The lower bound of $\mathbf{J}^D_{\mathbf{y}_\mathcal{Q}^{R/I}}(\tilde{\mathbf{h'}})$ is calculated based on the first and second order moments as
\begin{equation}
\mathbf{J}^D_{\mathbf{y}_\mathcal{Q}^{R/I}}(\tilde{\mathbf{h'}})\geq\left(\frac{\partial \mathbf{\mu}_{\mathbf{y}_\mathcal{Q}^{R/I}}}{\partial \tilde{\mathbf{h'}}}\right)^T\mathbf{C}^{-1}_{\mathbf{y}^{R/I}_\mathcal{Q}}\left(\frac{\partial \mathbf{\mu}_{\mathbf{y}_\mathcal{Q}^{R/I}}}{\partial \tilde{\mathbf{h'}}}\right) = \tilde{\mathbf{J}}^D_{\mathbf{y}_Q^{R/I}}(\tilde{\mathbf{h'}}).
\end{equation}
Since the lower-bounding technique is identical for the real and the imaginary parts, only the derivation of $\tilde{\mathbf{J}}^D_{\mathbf{y}_\mathcal{Q}^R}(\tilde{\mathbf{h'}})$ is presented. The mean value of the $k$th received symbol is
\begin{equation}
\begin{aligned}
[\mathbf{\mu}_{\mathbf{y}_\mathcal{Q}^R}]_k &= \frac{1}{\sqrt{2}}p([\mathbf{y}_\mathcal{Q}]_k=+1\mid\tilde{\mathbf{h'}})-\frac{1}{\sqrt{2}}p([\mathbf{y}_\mathcal{Q}]_k=-1\mid\tilde{\mathbf{h'}})\\&=\frac{1}{\sqrt{2}}\left[1-2Q\left(\frac{[\mathbf{\Phi}^R\mathbf{h'}^R-\mathbf{\Phi}^I\mathbf{h'}^I]_k}{\sqrt{[\mathbf{C}_\mathbf{n}]_{kk}/2}}\right)\right].
\end{aligned}
\label{equ_miu}
\end{equation}
The partial derivative of (\ref{equ_miu}) with respect to $[\tilde{\mathbf{h'}}]_i$ is
\begin{equation}
\frac{\partial [\mathbf{\mu}_{\mathbf{y}_\mathcal{Q}^R}]_k}{\partial [\tilde{\mathbf{h'}}]_i}=\frac{2\text{exp}\left(-\frac{[\mathbf{\Phi}^R\mathbf{h'}^R-\mathbf{\Phi}^I\mathbf{h'}^I]_k^2}{[\mathbf{C}_n]_{kk}}\right)\frac{\partial [\mathbf{\Phi}^R\mathbf{h'}^R-\mathbf{\Phi}^I\mathbf{h'}^I]_k}{\partial [\tilde{\mathbf{h'}}]_i}}{\sqrt{2\pi[\mathbf{C}_\mathbf{n}]_{kk}}}.
\end{equation}
The diagonal elements of the covariance matrix are given by
\begin{equation}
[\mathbf{C}_{\mathbf{y}^R_\mathcal{Q}}]_{kk}=\frac{1}{2}-[\mathbf{\mu}_{\mathbf{y}_\mathcal{Q}^R}]_k^2,
\end{equation}
while the off-diagonal elements are calculated as
\begin{equation}
\begin{aligned}
[\mathbf{C}_{\mathbf{y}^R_\mathcal{Q}}]_{kn}&=p(z_k>0,z_n>0)+p(z_k\leq0,z_n\leq0)\\&\hspace{2.5cm}-\frac{1}{2}-[\mathbf{\mu}_{\mathbf{y}_\mathcal{Q}^R}]_k[\mathbf{\mu}_{\mathbf{y}_\mathcal{Q}^R}]_n,
\end{aligned}
\end{equation}
where $[z_k,z_n]^T$ is a bi-variate Gaussian random vector
\begin{equation*}
\resizebox{\columnwidth}{!}{$\displaystyle
\begin{bmatrix}
z_k\\z_n
\end{bmatrix}\sim\mathcal{N}\left(\begin{bmatrix}
[\mathbf{\Phi}^R\mathbf{h'}^R-\mathbf{\Phi}^I\mathbf{h'}^I]_k\\ [\mathbf{\Phi}^R\mathbf{h'}^R-\mathbf{\Phi}^I\mathbf{h'}^I]_n
\end{bmatrix},\frac{1}{2}\begin{bmatrix}
[\mathbf{C}_\mathbf{n}]_{kk} & [\mathbf{C}_\mathbf{n}]_{kn}\\
[\mathbf{C}_\mathbf{n}]_{nk} & [\mathbf{C}_\mathbf{n}]_{nn}
\end{bmatrix}\right).$}
\end{equation*}
The lower bound for the imaginary part is derived in the same way. With the calculations above the lower bound of the BIM is obtained as
\begin{equation}
\mathbf{J}_{\mathbf{y}_\mathcal{Q}}(\tilde{\mathbf{h'}})\geq \tilde{\mathbf{J}}^D_{\mathbf{y}_\mathcal{Q}}(\tilde{\mathbf{h'}})+\mathbf{J}^P_{\mathbf{y}_\mathcal{Q}}(\tilde{\mathbf{h'}}),
\end{equation}
where the equality holds for $M=1$, as shown in \cite{Stein} for the real valued CRB and in \cite{Schluter} for the complex valued CRB. Based on (\ref{equ_bounds}), the inverse of this BIM lower bound will result in an upper bound of the actual Bayesian CRB for oversampled systems.

\subsection{General Cram\'er-Rao Bounds}
When $\mathbf{R}_\mathbf{h'}$ is unknown and needs to be estimated at the receiver, the Bayesian CRBs will not be applicable. The general CRBs are derived for the proposed channel estimators with estimated $\mathbf{R}_\mathbf{h'}$.
\begin{lemma}
    The proposed LRA channel estimators with combination of estimated $\hat{\mathbf{R}}_\mathbf{h'}$ are biased channel estimators.
\end{lemma}
\begin{IEEEproof}
    See Appendix C.
\end{IEEEproof}

Since the proposed LRA channel estimators are biased, while calculating the CRBs, they should apply as
\begin{equation}
Cov\{\hat{\mathbf{h'}}^R_{\text{bias}}\}\geq\frac{\partial E\{\hat{\mathbf{h'}}^R_{\text{bias}}\}}{\partial \mathbf{h'}^R}\left({\mathbf{J}^{D^{RR}}_{\mathbf{y}_\mathcal{Q}}(\mathbf{h'}^R)}^{-1}\frac{\partial E\{\hat{\mathbf{h'}}^R_{\text{bias}}\}}{\partial \mathbf{h'}^R}\right)^T
\label{equ_GCRB1}
\end{equation}
\begin{equation}
Cov\{\hat{\mathbf{h'}}^I_{\text{bias}}\}\geq\frac{\partial E\{\hat{\mathbf{h'}}^I_{\text{bias}}\}}{\partial \mathbf{h'}^I}\left(\mathbf{J}^{D^{II}}_{\mathbf{y}_\mathcal{Q}}(\mathbf{h'}^I)^{-1}\frac{\partial E\{\hat{\mathbf{h'}}^I_{\text{bias}}\}}{\partial \mathbf{h'}^I}\right)^T,
\label{equ_GCRB2}
\end{equation}
where $\mathbf{J}^{D^{RR}}_{\mathbf{y}_\mathcal{Q}}(\mathbf{h'}^R)$ and $\mathbf{J}^{D^{II}}_{\mathbf{y}_\mathcal{Q}}(\mathbf{h'}^I)$ are defined by
\begin{equation}
[\mathbf{J}^{D^{RR}}_{\mathbf{y}_\mathcal{Q}}(\mathbf{h'}^R)]_{ij}\triangleq E\left\{\frac{\partial \ln p(\mathbf{y}_\mathcal{Q}\mid\mathbf{h'}^R)}{\partial [\mathbf{h'}^R]_i}\frac{\partial \ln p(\mathbf{y}_\mathcal{Q}\mid\mathbf{h'}^R)}{\partial [\mathbf{h'}^R]_j}\right\}
\end{equation}
\begin{equation}
[\mathbf{J}^{D^{II}}_{\mathbf{y}_\mathcal{Q}}(\mathbf{h'}^I)]_{ij}\triangleq E\left\{\frac{\partial \ln p(\mathbf{y}_\mathcal{Q}\mid\mathbf{h'}^I)}{\partial [\mathbf{h'}^I]_i}\frac{\partial \ln p(\mathbf{y}_\mathcal{Q}\mid\mathbf{h'}^I)}{\partial [\mathbf{h'}^I]_j}\right\},
\end{equation}
which are the upper left and lower right part of the $\mathbf{J}^{D}_{\mathbf{y}_\mathcal{Q}}(\tilde{\mathbf{h'}})$ (similar as (\ref{equ_Jmatrix})), respectively.

\section{Numerical Results}
The simulation results presented here consider an uplink single-cell 1-bit large-scale MIMO system with $N_t = 8$ and $N_r = 64$. The modulation scheme is quadrature phase-shift keying (QPSK). The $m(t)$ and $p(t)$ filters are normalized RRC filters with a roll-off factor of 0.8. The channel is assumed to experience block fading and the pilots are column-wise orthogonal with length 20. The SNR is defined as $10\log(\frac{N_t}{\sigma_n^2})$. The normalized MSE and SER performance plots are obtained by taking the average of 300 channel matrices, noise and symbol vectors.

For the LRA-LMS channel estimator, the window length $l_{\text{win}}$ is chosen as three to ensure low computational complexity. The step size $\mu$ is optimized according to the oversampling factor and SNR. In the simulation, $\mu$ varies between 0.05 and 0.3. While recovering the transmitted symbols from the received quantized signal, the sliding-window based LMMSE detector \cite{Shao3} with window length equal to three ($l_{\text{win}}=3$) and the estimate of the channel obtained by the proposed algorithms is applied in the system for obtaining both high accuracy and low computational cost. We remark that more sophisticated detectors could also be considered \cite{mmimo,wence,mberdf,itic,spa,stmf,jiomimo,mfsic,wlmwf,dfcc,tdscl,tds,mbdf,did,rrmser,bfidd,1bitcpm,1bitidd,aaidd} along with precoders \cite{lclrbd,gbd,wlbd,mbthp,rmbthp,bbprec}.

The performance of the channel estimators is evaluated based on the channel model simulated in \cite{Clerckx}. The channel for user $n_t$ is assumed Rayleigh distributed
\begin{equation}
\mathbf{h'}_{n_t} = \mathbf{R}_{r,n_t}^{\frac{1}{2}}\mathbf{h}'_{w,n_t},
\end{equation}
where $\mathbf{R}_{r,n_t}$ denotes the receive correlation matrix with the following form
\begin{equation}
\mathbf{R}_{r,n_t} = \begin{bmatrix}
1 & \rho_{n_t} & \dots & \rho_{n_t}^{(N_r-1)}\\
\rho_{n_t}^* & 1 & \dots & \rho_{n_t}^{(N_r-2)}\\
\vdots & \vdots & \ddots & \vdots\\
\rho_{n_t}^{*(N_r-1)} & \rho_{n_t}^{*(N_r-2)} & \dots & 1\\
\end{bmatrix}.
\end{equation}
$\rho_{n_t}$ is the correlation index of neighboring antennas. ($|\rho_{n_t}|$ = 0 represents an uncorrelated scenario and $|\rho_{n_t}|$ = 1 implies a fully correlated scenario.) The elements of $\mathbf{h}'_{w,n_t}$ are i.i.d. complex Gaussian random variables with zero mean and unit variance. All users are assumed to experience the same value of $|\rho_{n_t}|=|\rho|$ but different phases uniformly distributed over $2\pi$. The overall channel model is summarized as
\begin{equation}
    \mathbf{H'} = [\mathbf{h'}_1,\mathbf{h'}_2, \cdots, \mathbf{h'}_{n_t}]
\end{equation}
and $\mathbf{R}_\mathbf{h'}$ is calculated as
\begin{equation}
    \mathbf{R}_\mathbf{h'} = \begin{bmatrix}
    \mathbf{R}_{r,1} & \mathbf{0} & \dots & \mathbf{0}\\
    \mathbf{0} & \mathbf{R}_{r,2} & \dots & \mathbf{0}\\
    \vdots & \vdots & \ddots & \vdots\\
    \mathbf{0} & \mathbf{0} & \dots & \mathbf{R}_{r,n_t}\\
    \end{bmatrix}.
    \label{equ_Rh'}
\end{equation}

\subsection{$\mathbf{R}_\MakeLowercase{\mathbf{h'}}$ is known at the receiver}
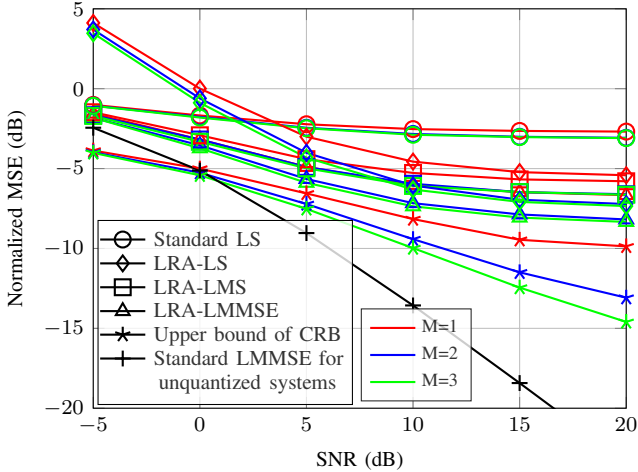
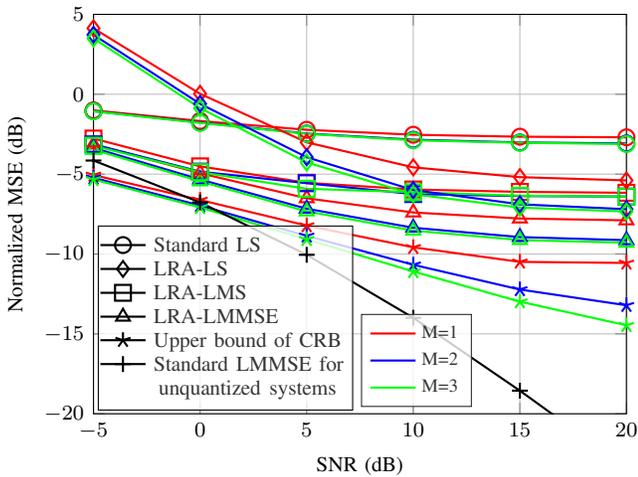
\begin{figure}[!htbp]
    \centering
    \begin{subfigure}{\columnwidth}
        \hspace{-0.8cm}
        \pgfplotsset{every axis label/.append style={font=\footnotesize},
	every tick label/.append style={font=\footnotesize},
}

\begin{tikzpicture}

\begin{axis}[%
width=.8\columnwidth,
height=.6\columnwidth,
at={(0.758in,0.603in)},
scale only axis,
xmin=-5,
xmax=20,
xlabel style={font=\footnotesize},
xlabel={SNR (dB)},
xtick=data,
xmajorgrids,
ymin=-20,
ymax=5,
yminorticks=true,
ylabel style={font=\footnotesize},
ylabel={Normalized MSE (dB)},
ymajorgrids,
yminorgrids,
axis background/.style={fill=white},
legend entries={M=1,
	M=2,
	M=3},
legend style={at={(0.71,0.25)},anchor=north east,legend cell align=left,align=left,draw=white!15!black,font=\scriptsize,fill opacity=0.8, draw opacity=1,text opacity=1}
]

\addlegendimage{color=red,fill=gray!20,line width=0.8pt,mark size=4pt}
\addlegendimage{color=blue,fill=green!20,line width=0.8pt,mark size=4pt}
\addlegendimage{color=green,fill=gray!20,line width=0.8pt,mark size=4pt}

\addplot [color=red,solid,line width=0.8pt,mark=triangle,mark options={solid},mark size=3pt]
table[row sep=crcr]{%
	-5  -1.56193102389158\\
	0	-3.17958048378023\\
	5   -4.89703009813506\\
	10	-6.02143932322647\\
	15	-6.47733694669436\\
	20  -6.66350240619888\\
};

\addplot [color=blue,solid,line width=0.8pt,mark=triangle,mark options={solid},mark size=3pt]
table[row sep=crcr]{%
	-5  -1.82280129863879\\
	0   -3.54632521064398\\
	5	-5.65126672243463\\
	10	-7.16522856206239\\
	15  -7.87106106194412\\
	20  -8.18454503468383\\
};

\addplot [color=green,solid,line width=0.8pt,mark=triangle,mark options={solid},mark size=3pt]
table[row sep=crcr]{%
	-5  -1.81725216631896\\  
	0	-3.72959003521146\\
	5	-5.89047082859027\\
	10	-7.36342766566363\\
	15	-8.06647768112484\\
	20	-8.33941305156611\\
};

\addplot [color=red,solid,line width=0.8pt,mark=square,mark options={solid},mark size=3pt]
table[row sep=crcr]{%
	-5  -1.4742\\
	0	-2.91469464699648\\
	5	-4.40936865089677\\
	10	-5.27409324201509\\
	15	-5.67784197565453\\
	20  -5.79390590677543\\
};

\addplot [color=blue,solid,line width=0.8pt,mark=square,mark options={solid},mark size=3pt]
table[row sep=crcr]{%
	-5   -1.6569\\
	0   -3.17957588259261\\
	5	-4.90093814108066\\
	10	-5.95087936621141\\
	15  -6.47505296187813\\
	20  -6.60425036714094\\
};

\addplot [color=green,solid,line width=0.8pt,mark=square,mark options={solid},mark size=3pt]
table[row sep=crcr]{%
	-5  -1.7054\\
	0	-3.32038954552652\\
	5	-4.98269668227407\\
	10	-6.08760819722245\\
	15	-6.47035258892312\\
	20	-6.63708213917238\\
};

\addplot [color=red,solid,line width=0.8pt,mark=o,mark options={solid},mark size=3pt]
table[row sep=crcr]{%
	-5  -1.01275593663253\\
	0   -1.67672792789365\\
	5	-2.22508241107315\\
	10	-2.52823917045215\\
	15	-2.64294487527566\\
	20  -2.68566411286690\\
};

\addplot [color=blue,solid,line width=0.8pt,mark=o,mark options={solid},mark size=3pt]
table[row sep=crcr]{%
	-5  -1.06150847889434\\
	0   -1.76558411197454\\
	5	-2.44473245506339\\
	10	-2.83330732838154\\
	15  -3.00968239306343\\
	20  -3.07373273633643\\
};

\addplot [color=green,solid,line width=0.8pt,mark=o,mark options={solid},mark size=3pt]
table[row sep=crcr]{%
	-5  -1.07147408893779\\
	0	-1.80694147840457\\
	5	-2.47427453646174\\
	10	-2.88389360464622\\
	15	-3.03009409268943\\
	20	-3.09304581755759\\
};

\addplot [color=red,solid,line width=0.8pt,mark=diamond,mark options={solid},mark size=3pt]
table[row sep=crcr]{%
	-5  4.11496567309527\\
	0   0.00648040732168850\\
	5	-3.00254862390456\\
	10	-4.56532093374834\\
	15	-5.22252523881746\\
	20  -5.41740829838311\\
};

\addplot [color=blue,solid,line width=0.8pt,mark=diamond,mark options={solid},mark size=3pt]
table[row sep=crcr]{%
	-5  3.70236222723397\\
	0   -0.605252048320847\\
	5	-4.01260200724474\\
	10	-6.06064407881082\\
	15  -6.94922709788979\\
	20  -7.22127965167770\\
};

\addplot [color=green,solid,line width=0.8pt,mark=diamond,mark options={solid},mark size=3pt]
table[row sep=crcr]{%
	-5  3.46819942144475\\
	0	-0.884311953398404\\
	5	-4.31092122313288\\
	10	-6.31062609264908\\
	15	-7.0860\\
	20	-7.3479\\
};

\addplot [color=red,solid,line width=0.8pt,mark=star,mark options={solid},mark size=3pt]
table[row sep=crcr]{%
	-5  -3.8748\\
	0	-5.00262920641373\\
	5	-6.55551772216329\\
	10	-8.16109265388428\\
	15	-9.45741915296547\\
	20  -9.86904334088659\\
};

\addplot [color=blue,solid,line width=0.8pt,mark=star,mark options={solid},mark size=3pt]
table[row sep=crcr]{%
	-5  -3.9794\\
	0	-5.25933445112959\\
	5	-7.21754227567162\\
	10	-9.41779366124783\\
	15  -11.4931676506731\\
	20	-13.0836156874015\\
};

\addplot [color=green,solid,line width=0.8pt,mark=star,mark options={solid},mark size=3pt]
table[row sep=crcr]{%
	-5  -4.0171\\
	0	-5.42100358115503\\
	5	-7.52163013106677\\
	10	-9.97599867731766\\
	15  -12.4491089466074\\
	20	-14.5983740781336\\
};

\addplot [color=black,solid,line width=0.8pt,mark=+,mark options={solid},mark size=3pt]
table[row sep=crcr]{%
	-5  -2.44340939388406\\
	0	-5.12204798927135\\
	5	-9.03615240702583\\
	10	-13.5658166350107\\
	15	-18.4242426817944\\
	20	-23.3715180066214\\
};

\addplot[only marks,smooth,color=black,solid,line width=0.8pt,mark=square,mark size=3pt,
y filter/.code={\pgfmathparse{\pgfmathresult+6}\pgfmathresult}]
table[row sep=crcr]{%
	1 2 3 4 5 6\\
};\label{L21}

\addplot[only marks,smooth,color=black,solid,line width=0.8pt,mark=triangle,mark size=3pt,y filter/.code={\pgfmathparse{\pgfmathresult+6}\pgfmathresult}]
table[row sep=crcr]{%
	1 2 3 4 5 6\\
};\label{L22}

\addplot[only marks,smooth,color=black,solid,line width=0.8pt,mark=o,mark size=3pt,y filter/.code={\pgfmathparse{\pgfmathresult+6}\pgfmathresult}]
table[row sep=crcr]{%
	1 2 3 4 5\\
};\label{L23}

\addplot[only marks,smooth,color=black,solid,line width=0.8pt,mark=diamond,mark size=3pt,y filter/.code={\pgfmathparse{\pgfmathresult+6}\pgfmathresult}]
table[row sep=crcr]{%
	1 2 3 4 5\\
};\label{L24}

\addplot[smooth,color=black,solid,line width=0.8pt,
y filter/.code={\pgfmathparse{\pgfmathresult+6}\pgfmathresult}]
table[row sep=crcr]{%
	1 2 3 4 5\\
};\label{L22221}

\addplot[smooth,color=black,dashed,line width=0.8pt,
y filter/.code={\pgfmathparse{\pgfmathresult+6}\pgfmathresult}]
table[row sep=crcr]{%
	1 2 3 4 5\\
};\label{L22222}

\addplot[only marks,smooth,color=black,solid,line width=0.8pt,mark=star,mark size=3pt,y filter/.code={\pgfmathparse{\pgfmathresult+6}\pgfmathresult}]
table[row sep=crcr]{%
	1 2\\
};\label{L25}

\addplot[only marks,smooth,color=black,solid,line width=0.8pt,mark=+,mark size=3pt,y filter/.code={\pgfmathparse{\pgfmathresult+6}\pgfmathresult}]
table[row sep=crcr]{%
	1 2\\
};\label{L26}

\node [draw,fill=white,font=\footnotesize,anchor= south west,at={(-4.8,-19.8)},fill opacity=0.8, draw opacity=1,text opacity=1] {
	\setlength{\tabcolsep}{0mm}
	\renewcommand{\arraystretch}{1}
	\begin{tabular}{l}
	\hspace{-0.3cm}\ref{L23}{~Standard LS}\\
	\hspace{-0.3cm}\ref{L24}{~LRA-LS}\\
	\hspace{-0.3cm}\ref{L21}{~LRA-LMS}\\
	\hspace{-0.3cm}\ref{L22}{~LRA-LMMSE}\\
	\hspace{-0.3cm}\ref{L25}{~Upper bound of CRB}\\
	\hspace{-0.3cm}\ref{L26}{~Standard LMMSE for}\\
	{~~~~~unquantized systems}\\
	\end{tabular}
};

\end{axis}

\end{tikzpicture}%
        \caption{$|\rho|=0$}
        \label{fig:MSE_com}
    \end{subfigure}
    \\
    \begin{subfigure}{\columnwidth}
        \hspace{-0.8cm}
        \pgfplotsset{every axis label/.append style={font=\footnotesize},
	every tick label/.append style={font=\footnotesize},
}

\begin{tikzpicture}

\begin{axis}[%
width=.8\columnwidth,
height=.6\columnwidth,
at={(0.758in,0.603in)},
scale only axis,
xmin=-5,
xmax=20,
xlabel style={font=\footnotesize},
xlabel={SNR (dB)},
xtick=data,
xmajorgrids,
ymin=-20,
ymax=5,
yminorticks=true,
ylabel style={font=\footnotesize},
ylabel={Normalized MSE (dB)},
ymajorgrids,
yminorgrids,
axis background/.style={fill=white},
legend entries={M=1,
	M=2,
	M=3},
legend style={at={(0.71,0.25)},anchor=north east,legend cell align=left,align=left,draw=white!15!black,font=\scriptsize,fill opacity=0.8, draw opacity=1,text opacity=1}
]

\addlegendimage{color=red,fill=gray!20,line width=0.8pt,mark size=4pt}
\addlegendimage{color=blue,fill=green!20,line width=0.8pt,mark size=4pt}
\addlegendimage{color=green,fill=gray!20,line width=0.8pt,mark size=4pt}


\addplot [color=red,solid,line width=0.8pt,mark=triangle,mark options={solid},mark size=3pt]
table[row sep=crcr]{%
	-5  -3.14509710007132\\
	0	-4.94283982138581\\
	5	-6.51123781599668\\
	10	-7.40030374962096\\
	15	-7.77676279081630\\
	20  -7.89247750214781\\
};

\addplot [color=blue,solid,line width=0.8pt,mark=triangle,mark options={solid},mark size=3pt]
table[row sep=crcr]{%
	-5  -3.35549888840733\\
	0   -5.33538002657551\\
	5	-7.16282615316021\\
	10	-8.35474283942393\\
	15  -8.94023261247215\\
	20  -9.13530895638639\\
};

\addplot [color=green,solid,line width=0.8pt,mark=triangle,mark options={solid},mark size=3pt]
table[row sep=crcr]{%
	-5  -3.46386964499272\\
	0	-5.45280083207966\\
	5	-7.33393749180654\\
	10	-8.53604284689617\\
	15	-9.12480960376379\\
	20	-9.29307879098834\\
};


\addplot [color=red,solid,line width=0.8pt,mark=square,mark options={solid},mark size=3pt]
table[row sep=crcr]{%
	-5  -2.7713\\
	0	-4.5131\\
	5	-5.5415\\
	10	-5.9611\\
	15  -6.1093\\
	20  -6.16783623313995\\
};

\addplot [color=blue,solid,line width=0.8pt,mark=square,mark options={solid},mark size=3pt]
table[row sep=crcr]{%
	-5  -3.1329\\ 
	0   -4.8434\\
	5	-5.5690\\
	10	-6.2479\\
	15  -6.3709\\
	20  -6.4045\\
};

\addplot [color=green,solid,line width=0.8pt,mark=square,mark options={solid},mark size=3pt]
table[row sep=crcr]{%
	-5  -3.2226\\
	0	-4.8769\\
	5	-5.9164\\
	10	-6.1329\\
	15	-6.3466\\
	20  -6.4019\\
};


\addplot [color=red,solid,line width=0.8pt,mark=o,mark options={solid},mark size=3pt]
table[row sep=crcr]{%
	-5  -1.00850663177467\\
	0	-1.67345340061805\\
	5	-2.22480480629636\\
	10  -2.53491118403009\\
	15  -2.65731089350277\\
	20  -2.69608714299894\\
};

\addplot [color=blue,solid,line width=0.8pt,mark=o,mark options={solid},mark size=3pt]
table[row sep=crcr]{%
	-5  -1.05957364493234\\
	0   -1.76950561943346\\
	5	-2.45845440099932\\
	10	-2.84462923935511\\
	15  -3.01837033973831\\
	20  -3.07645177161810\\
};

\addplot [color=green,solid,line width=0.8pt,mark=o,mark options={solid},mark size=3pt]
table[row sep=crcr]{%
	-5  -1.06898293012350\\
	0	-1.79460557450190\\
	5	-2.48105138727322\\
	10	-2.88318852647559\\
	15	-3.03893342444317\\
	20  -3.11687510578905\\
};

\addplot [color=red,solid,line width=0.8pt,mark=diamond,mark options={solid},mark size=3pt]
table[row sep=crcr]{%
	-5  4.12714374747439\\
	0   0.0286741527703726\\
	5	-3.00856276311527\\
	10  -4.57468434152370\\
	15  -5.18833943636315\\
	20  -5.39322917304424\\
};

\addplot [color=blue,solid,line width=0.8pt,mark=diamond,mark options={solid},mark size=3pt]
table[row sep=crcr]{%
	-5  3.72347847915975\\
	0   -0.598868624364578\\
	5	-3.94039141276611\\
	10	-5.99542061245864\\
	15  -6.88865444820006\\
	20  -7.19510744470983\\
};

\addplot [color=green,solid,line width=0.8pt,mark=diamond,mark options={solid},mark size=3pt]
table[row sep=crcr]{%
	-5   3.49756436427933\\
	0   -0.866983961529285\\
	5	-4.26711713876165\\
	10	-6.26024969241295\\
	15	-7.09503535114325\\
	20  -7.34117497938652\\
};

\addplot [color=red,solid,line width=0.8pt,mark=star,mark options={solid},mark size=3pt]
table[row sep=crcr]{%
	-10 -3.9133\\
	-5  -5.0556\\
	0	-6.61787654129567\\
	5	-8.21996042914690\\
	10	-9.57985217996811\\
	15  -10.4918561429403\\
	20  -10.5593088948746\\
};

\addplot [color=blue,solid,line width=0.8pt,mark=star,mark options={solid},mark size=3pt]
table[row sep=crcr]{%
	-10 -3.99193236358634\\
	-5  -5.22534578660916\\
	0	-6.96555068513797\\
	5	-8.86528392064944\\
	10	-10.6734253349557\\
	15  -12.2224310758635\\
	20	-13.2080797953817\\
};

\addplot [color=green,solid,line width=0.8pt,mark=star,mark options={solid},mark size=3pt]
table[row sep=crcr]{%
	-10 -4.0593\\
	-5  -5.3337\\
	0	-7.0822\\
	5	-9.0851\\
	10	-11.0915\\
	15	-12.9810\\
	20	-14.4615\\
};

\addplot [color=black,solid,line width=0.8pt,mark=+,mark options={solid},mark size=3pt]
table[row sep=crcr]{%
	-5  -4.14825012111980\\
	0	-6.76742678220672\\
	5	-10.0488842127520\\
	10	-13.9913955062152\\
	15  -18.5544693381498\\
	20	-23.3710430571636\\
};

\node [draw,fill=white,font=\footnotesize,anchor= south west,at={(-4.8,-19.8)},,fill opacity=0.8, draw opacity=1,text opacity=1] {
	\setlength{\tabcolsep}{0mm}
	\renewcommand{\arraystretch}{1}
	\begin{tabular}{l}
	\hspace{-0.3cm}\ref{L23}{~Standard LS}\\
	\hspace{-0.3cm}\ref{L24}{~LRA-LS}\\
	\hspace{-0.3cm}\ref{L21}{~LRA-LMS}\\
	\hspace{-0.3cm}\ref{L22}{~LRA-LMMSE}\\
	\hspace{-0.3cm}\ref{L25}{~Upper bound of CRB}\\
	\hspace{-0.3cm}\ref{L26}{~Standard LMMSE for}\\
	{~~~~~unquantized systems}\\
	\end{tabular}
};

\end{axis}

\end{tikzpicture}%
        \caption{$|\rho|=0.75$}
        \label{fig:MSE_com2}
    \end{subfigure}
    \caption{Normalized MSE comparisons of different channel estimators with known $\mathbf{R}_\mathbf{h'}$}
    \label{fig_MSE_t}
\end{figure}

In this subsection, we evaluate the performance of the proposed LRA channel estimators with known $\mathbf{R}_\mathbf{h'}$ at the receiver. Fig. \ref{fig:MSE_com} and Fig. \ref{fig:MSE_com2} compare the normalized MSE of the various channel estimators as a function of SNR in uncorrelated ($|\rho|=0$) and correlated channel ($|\rho|=0.75$), respectively. There is a 2dB performance gain of the oversampled systems as compared to the non-oversampled systems for the LRA-LMMSE channel estimator at low SNR, whereas a much larger gain at high SNR. In both channels the LRA-LMMSE achieves the best MSE performance at the cost of high computational cost.

In contrast, the LRA-LMS estimates the channel matrix $\mathbf{H'}$ row by row. This approach can largely reduce the computational cost (shown in Fig. \ref{fig:Complexity}). Note that this separation into several rows may overlook the correlation of receive antennas. More specifically, the proposed LRA-LMS treats $\mathbf{R}_{r,n_t}$ as an identity matrix. As an amendment, the resulting estimated channel matrix $\hat{\mathbf{h}}'_{\text{LRA-LMS}}$ needs to be multiplied with the square root of the receive correlation matrix $\mathbf{R}_{r_{n_t}}^{\frac{1}{2}}$, which can be derived from $\mathbf{R}_\mathbf{h'}$ in (\ref{equ_Rh'}). From the results, it can be seen that in both channels the LRA-LMS approaches the performance of the LRA-LMMSE at low SNR ($\leq 5$ dB), whereas at high SNR this performance gap becomes large.

The Bayesian CRBs illustrated in Section \Romannum{4}-A are also depicted in Fig. \ref{fig_MSE_t}. Note that for the oversampled systems ($M\geq2$) the upper bounds of Bayesian CRBs are higher than the actual Bayesian CRBs, since they are derived from the lower bounds of Bayesian information. The black lines represent the standard LMMSE performance for the systems with unquantized signals, which can be treated as lower bounds for the systems with 1-bit quantized signals.


%


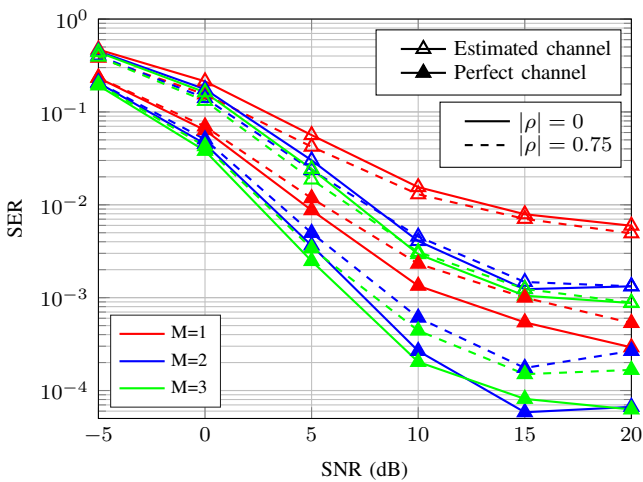
\begin{figure}[!htbp]
    \centering
    \pgfplotsset{every axis label/.append style={font=\footnotesize},
	every tick label/.append style={font=\footnotesize},
}

\begin{tikzpicture}

\begin{axis}[%
width=.8\columnwidth,
height=.6\columnwidth,
at={(0.758in,0.603in)},
scale only axis,
xmin=-5,
xmax=20,
xlabel style={font=\footnotesize},
xlabel={SNR (dB)},
xtick=data,
xmajorgrids,
ymode=log,
ymin=5e-5,
ymax=1,
yminorticks=true,
ylabel style={font=\footnotesize},
ylabel={SER},
ymajorgrids,
yminorgrids,
axis background/.style={fill=white},
legend entries={M=1,
	M=2,
	M=3},
legend style={at={(0.23,0.26)},anchor=north east,legend cell align=left,align=left,draw=white!15!black,font=\scriptsize}
]

\addlegendimage{color=red,fill=gray!20,line width=0.8pt,mark size=3pt}
\addlegendimage{color=blue,fill=green!20,line width=0.8pt,mark size=3pt}
\addlegendimage{color=green,fill=gray!20,line width=0.8pt,mark size=3pt}

\addplot [color=red,solid,line width=0.8pt,mark=triangle,mark options={solid},mark size=3pt]
table[row sep=crcr]{%
	-5  0.469125000000000\\
	0	0.213525000000000\\
	5	0.0562750000000000\\ 
	10	0.0154833333333333\\
	15  0.00790833333333331\\
	20	0.00594999999999999\\
};

\addplot [color=red,dashed,line width=0.8pt,mark=triangle,mark options={solid},mark size=3pt]table[row sep=crcr]{%
	-5  0.3830\\
	0	0.154416666666667\\
	5	0.0421750000000000\\ 
	10	0.0129666666666666\\
	15	0.00704999999999998\\
	20	0.00492499999999998\\
};

\addplot [color=blue,solid,line width=0.8pt,mark=triangle,mark options={solid},mark size=3pt]table[row sep=crcr]{%
	-5  0.448783333333333\\
	0	0.175858333333333\\
	5	0.0301250000000000\\ 
	10	0.00406666666666665\\
	15	0.00123333333333333\\
	20	0.00132500000000000\\
};

\addplot [color=blue,dashed,line width=0.8pt,mark=triangle,mark options={solid},mark size=3pt]table[row sep=crcr]{%
	-5  0.4035\\
	0	0.141075000000000\\
	5   0.0236250000000000\\ 
	10	0.00454166666666665\\
	15	0.00147500000000000\\
	20	0.00132500000000000\\
};

\addplot [color=green,solid,line width=0.8pt,mark=triangle,mark options={solid},mark size=3pt]table[row sep=crcr]{%
	-5  0.439741666666667\\
	0	0.162783333333333\\
	5	0.0243000000000000\\ 
	10	0.00294166666666666\\
	15	0.00105000000000000\\
	20	0.000875000000000001\\
};

\addplot [color=green,dashed,line width=0.8pt,mark=triangle,mark options={solid},mark size=3pt]
table[row sep=crcr]{%
	-5  0.3934\\
	0   0.131241666666667\\
	5	0.0187000000000000\\ 
	10	0.00309999999999999\\
	15	0.00124166666666667\\
	20	0.000883333333333334\\ 
};

\addplot [color=red,solid,line width=0.8pt,mark=triangle*,mark options={solid},mark size=3pt]
table[row sep=crcr]{%
	-5  0.232616666666667\\
	0   0.0636250000000001\\
	5	0.00873333333333331\\ 
	10	0.00134166666666667\\
	15	0.000541666666666667\\
	20	0.000291666666666667\\
};

\addplot [color=red,dashed,line width=0.8pt,mark=triangle*,mark options={solid},mark size=3pt]
table[row sep=crcr]{%
	-5  0.234083333333333\\
	0   0.0697250000000000\\
	5   0.0117666666666666\\ 
	10  0.00231666666666666\\
	15	0.00100000000000000\\
	20	0.000533333333333334\\
};

\addplot [color=blue,solid,line width=0.8pt,mark=triangle*,mark options={solid},mark size=3pt]
table[row sep=crcr]{%
	-5  0.205816666666667\\
	0	0.0452083333333333\\
	5	0.00364166666666665\\ 
	10  0.000266666666666667\\
	15	0.0000583333333333333\\
	20	0.0000666666666666667\\
};

\addplot [color=blue,dashed,line width=0.8pt,mark=triangle*,mark options={solid},mark size=3pt]
table[row sep=crcr]{%
	-5  0.207850000000000\\
	0   0.0506083333333333\\
	5	0.00497499999999998\\ 
	10	0.000608333333333334\\
	15	0.000175000000000000\\
	20  0.000266666666666667\\
};

\addplot [color=green,solid,line width=0.8pt,mark=triangle*,mark options={solid},mark size=3pt]
table[row sep=crcr]{%
	-5  0.192675000000000\\
	0	0.0380750000000000\\
	5	0.00247812499999998\\ 
	10	0.000203125000000000\\
	15	0.00008125\\
	20	0.0000625\\
};

\addplot [color=green,dashed,line width=0.8pt,mark=triangle*,mark options={solid},mark size=3pt]
table[row sep=crcr]{%
	-5  0.198850000000000\\
	0	0.0427416666666666\\
	5	0.00339999999999999\\ 
	10	0.000441666666666667\\
	15	0.000150000000000000\\
	20	0.000166666666666667\\ 
};

\addplot[only marks,smooth,color=black,solid,line width=0.8pt,mark=triangle,mark size=3pt,
y filter/.code={\pgfmathparse{\pgfmathresult-0}\pgfmathresult}]
table[row sep=crcr]{%
	1 2\\
};\label{L2221}

\addplot[only marks,smooth,color=black,solid,line width=0.8pt,mark=triangle*,mark size=3pt,
y filter/.code={\pgfmathparse{\pgfmathresult-0}\pgfmathresult}]
table[row sep=crcr]{%
	1 2\\
};\label{L2222}

\node [draw,fill=white,font=\footnotesize,anchor= south west,at={(8,0.15)}] {
	\setlength{\tabcolsep}{0mm}
	\renewcommand{\arraystretch}{1}
	\begin{tabular}{l}
	\ref{L2221}{~Estimated channel}\\
	\ref{L2222}{~Perfect channel}\\
	\end{tabular}
};

\node [draw,fill=white,font=\footnotesize,anchor= south west,at={(11,0.025)}] {
	\setlength{\tabcolsep}{0mm}
	\renewcommand{\arraystretch}{1}
	\begin{tabular}{l}
	\ref{L22221}{~$|\rho|=0$}\\
	\ref{L22222}{~$|\rho|=0.75$}\\
	\end{tabular}
};

\end{axis}
\end{tikzpicture}%
    \caption{SER comparisons of different oversampling factors for the LRA-LMMSE channel estimator with known $\mathbf{R}_\mathbf{h'}$.}
    \label{fig:SER}
\end{figure}

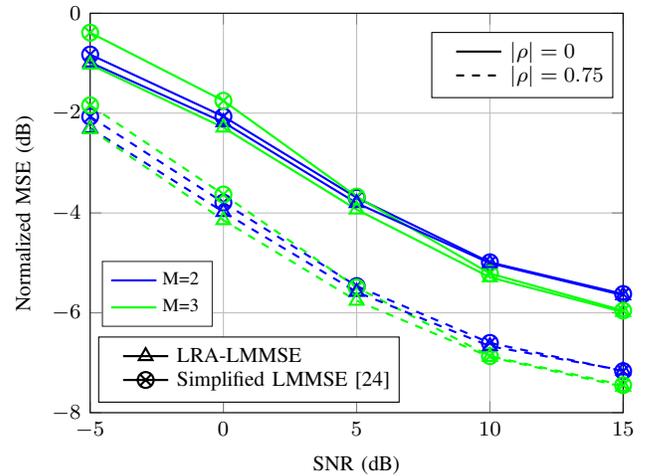
\begin{figure}[!htbp]
    \centering
    \pgfplotsset{every axis label/.append style={font=\footnotesize},
	every tick label/.append style={font=\footnotesize},
}

\begin{tikzpicture}

\begin{axis}[%
width=.8\columnwidth,
height=.6\columnwidth,
at={(0.758in,0.603in)},
scale only axis,
xmin=-5,
xmax=15,
xlabel style={font=\footnotesize},
xlabel={SNR (dB)},
xtick=data,
xmajorgrids,
ymin=-8,
ymax=0,
yminorticks=true,
ylabel style={font=\footnotesize},
ylabel={Normalized MSE (dB)},
ymajorgrids,
yminorgrids,
axis background/.style={fill=white},
legend entries={
	M=2,
	M=3},
legend style={at={(0.23,0.38)},anchor=north east,legend cell align=left,align=left,draw=white!15!black,font=\scriptsize}
]

\addlegendimage{color=blue,fill=green!20,line width=0.8pt,mark size=4pt}
\addlegendimage{color=green,fill=gray!20,line width=0.8pt,mark size=4pt}

\addplot [color=blue,solid,line width=0.8pt,mark=otimes,mark options={solid},mark size=3pt]
table[row sep=crcr]{%
	-5  -0.826546276257581\\
	0   -2.06601820726596\\
	5   -3.69370055006320\\
	10  -4.98783148496509\\
	15  -5.62252160297499\\
};

\addplot [color=green,solid,line width=0.8pt,mark=otimes,mark options={solid},mark size=3pt]
table[row sep=crcr]{%
	-5  -0.387608485302145\\
	0   -1.75229241939624\\
	5   -3.67118670988824\\
	10  -5.21187654175496\\
	15  -5.95627633191448\\
};

\addplot [color=blue,dashed,line width=0.8pt,mark=otimes,mark options={solid},mark size=3pt]
table[row sep=crcr]{%
	-5  -2.07026467464581\\
	0   -3.79242006254786\\
	5   -5.46435050912319\\
	10  -6.60213355773125\\
	15  -7.1666\\
};

\addplot [color=green,dashed,line width=0.8pt,mark=otimes,mark options={solid},mark size=3pt]
table[row sep=crcr]{%
	-5  -1.84316498811294\\
	0   -3.63056819082492\\
	5   -5.49164234883601\\
	10  -6.86754522052391\\
	15  -7.44865414876378\\
};

\addplot [color=blue,solid,line width=0.8pt,mark=triangle,mark options={solid},mark size=3pt]
table[row sep=crcr]{%
	-5  -0.985335857484078\\
	0   -2.18296050997266\\
	5   -3.80214532687454\\
	10  -5.00534535591195\\
	15  -5.64796022904642\\
};

\addplot [color=green,solid,line width=0.8pt,mark=triangle,mark options={solid},mark size=3pt]
table[row sep=crcr]{%
	-5  -1.02763388910387\\
	0   -2.28946386829363\\
	5   -3.93089332783803\\
	10  -5.28848094071469\\
	15  -5.98334248268231\\
};

\addplot [color=blue,dashed,line width=0.8pt,mark=triangle,mark options={solid},mark size=3pt]
table[row sep=crcr]{%
	-5  -2.30113006239005\\
	0   -3.98007365327890\\
	5   -5.58155662295991\\
	10  -6.66717879086188\\
	15  -7.15377073067993\\
};

\addplot [color=green,dashed,line width=0.8pt,mark=triangle,mark options={solid},mark size=3pt]
table[row sep=crcr]{%
	-5  -2.32136540325065\\
	0   -4.14426691852954\\
	5   -5.75794233121142\\
	10  -6.88329216833846\\
	15  -7.47887901660508\\
};

\addplot[only marks,smooth,color=black,solid,line width=0.8pt,mark=otimes,mark size=3pt,y filter/.code={\pgfmathparse{\pgfmathresult-0}\pgfmathresult}]
table[row sep=crcr]{%
	1 2\\
};\label{L28}

\node [draw,fill=white,font=\footnotesize,anchor= south west,at={(-4.7,-7.8)}] {
	\setlength{\tabcolsep}{0mm}
	\renewcommand{\arraystretch}{1}
	\begin{tabular}{l}
	\ref{L22}{~LRA-LMMSE}\\
	\ref{L28}{~Simplified LMMSE \cite{Ucuncu}}\\
	\end{tabular}
};

\node [draw,fill=white,font=\footnotesize,anchor= south west,at={(7.8,-1.7)}] {
	\setlength{\tabcolsep}{0mm}
	\renewcommand{\arraystretch}{1}
	\begin{tabular}{l}
	\ref{L22221}{~$|\rho|=0$}\\
	\ref{L22222}{~$|\rho|=0.75$}\\
	\end{tabular}
};

\end{axis}

\end{tikzpicture}%
    \caption{Normalized MSE comparisons between LRA-LMMSE and simplified LMMSE \cite{Ucuncu}.}
    \label{fig:MSE_Cn}
\end{figure}

The LMMSE detector with sliding-window based SER performance of the system with the LRA-LMMSE estimated and perfect channel matrix are illustrated in Fig. \ref{fig:SER}, where the oversampled systems obviously outperform the non-oversampled systems. As described in \Romannum{3}-A, Fig. \ref{fig:MSE_Cn} shows the MSE comparisons between LRA-LMMSE and simplified LMMSE \cite{Ucuncu} channel estimator in the system with $\tau=10$ and roll-off factor 0.1. We emphasize again that in our work, the correlation of filtered noise is taken into account, and hence $\mathbf{C}_{\mathbf{n}_p}$ is not a diagonal matrix in oversampled systems. It can be seen that at low SNR ($\leq$ 10 dB) the performance of simplified LMMSE \cite{Ucuncu} is worse than the proposed LRA-LMMSE, although they converge together at high SNR (> 10 dB). Another observation is that at low SNR the simplified LMMSE estimator with $M=3$ performs worse than that with $M=2$, which shows that the assumption in \cite{Ucuncu} is inaccurate.

\subsection{$\mathbf{R}_\MakeLowercase{\mathbf{h'}}$ is unknown at the receiver}
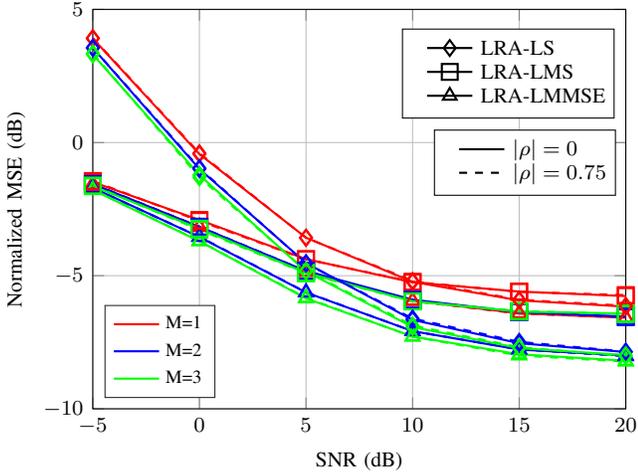
\begin{figure}[!htbp]
    \centering
    \pgfplotsset{every axis label/.append style={font=\footnotesize},
	every tick label/.append style={font=\footnotesize},
}

\begin{tikzpicture}

\begin{axis}[%
width=.8\columnwidth,
height=.6\columnwidth,
at={(0.758in,0.603in)},
scale only axis,
xmin=-5,
xmax=20,
xlabel style={font=\footnotesize},
xlabel={SNR (dB)},
xtick=data,
xmajorgrids,
ymin=-10,
ymax=5,
yminorticks=true,
ylabel style={font=\footnotesize},
ylabel={Normalized MSE (dB)},
ymajorgrids,
yminorgrids,
axis background/.style={fill=white},
legend entries={M=1,
	M=2,
	M=3},
legend style={at={(0.23,0.26)},anchor=north east,legend cell align=left,align=left,draw=white!15!black,font=\scriptsize}
]

\addlegendimage{color=red,fill=gray!20,line width=0.8pt,mark size=3pt}
\addlegendimage{color=blue,fill=green!20,line width=0.8pt,mark size=3pt}
\addlegendimage{color=green,fill=gray!20,line width=0.8pt,mark size=3pt}

\addplot [color=red,dashed,line width=0.8pt,mark=triangle,mark options={solid},mark size=3pt]
table[row sep=crcr]{
	-5  -1.56335194839808\\
	0	-3.17460084860423\\
	5	-4.85971957753317\\
	10	-5.95175211398891\\
	15	-6.42328933182422\\
	20  -6.56758083568983\\
};

\addplot [color=red,solid,line width=0.8pt,mark=triangle,mark options={solid},mark size=3pt]
table[row sep=crcr]{
	-5  -1.55926418505797\\
	0	-3.16442525694894\\
	5	-4.85285456603868\\
	10	-5.93634570614217\\
	15	-6.41230893569280\\
	20  -6.58020003443034\\
};

\addplot [color=blue,dashed,line width=0.8pt,mark=triangle,mark options={solid},mark size=3pt]
table[row sep=crcr]{
	-5  -1.71101342509830\\
	0   -3.53610129014181\\
	5	-5.62368236121925\\
	10	-7.08238655808942\\
	15  -7.76676274976952\\
	20  -8.02539896664991\\
};

\addplot [color=blue,solid,line width=0.8pt,mark=triangle,mark options={solid},mark size=3pt]
table[row sep=crcr]{
	-5  -1.69643912155629\\
	0	-3.54852198479639\\
	5	-5.63088594161681\\
	10	-7.08152251731024\\
	15	-7.77450962535104\\
	20  -7.99653113307437\\
};

\addplot [color=green,dashed,line width=0.8pt,mark=triangle,mark options={solid},mark size=3pt]
table[row sep=crcr]{
	-5  -1.76912482685999\\
	0	-3.69952633365483\\
	5	-5.83050253802198\\
	10	-7.29112536196821\\
	15	-7.99778372519560\\
	20	-8.22894836959579\\
};

\addplot [color=green,solid,line width=0.8pt,mark=triangle,mark options={solid},mark size=3pt]
table[row sep=crcr]{
	-5  -1.77885877106509\\
	0	-3.70619863238265\\
	5	-5.85366046505601\\
	10	-7.28166523857904\\
	15	-7.95234659197626\\
	20	-8.19472782798674\\
};

\addplot [color=red,dashed,line width=0.8pt,mark=square,mark options={solid},mark size=3pt]
table[row sep=crcr]{%
	-5  -1.44439884780542\\
	0	-2.94395413488487\\
	5	-4.38560710570838\\
	10	-5.22934120374687\\
	15	-5.60754268431478\\
	20  -5.72785731019185\\
};

\addplot [color=red,solid,line width=0.8pt,mark=square,mark options={solid},mark size=3pt]
table[row sep=crcr]{
	-5  -1.48389139293796\\
	0	-2.90763509201715\\
	5	-4.37627831757040\\
	10	-5.24929761146050\\
	15	-5.59686765033404\\
	20  -5.76583514241057\\
};

\addplot [color=blue,dashed,line width=0.8pt,mark=square,mark options={solid},mark size=3pt]
table[row sep=crcr]{
	-5  -1.55690104908062\\
	0   -3.15951796782733\\
	5	-4.82824844515723\\
	10	-5.94046067332182\\
	15  -6.37088631517243\\
	20  -6.55957267374586\\
};

\addplot [color=blue,solid,line width=0.8pt,mark=square,mark options={solid},mark size=3pt]
table[row sep=crcr]{
	-5  -1.56337024865127\\
	0   -3.17406015860785\\
	5	-4.82199799230869\\
	10  -5.90147402026571\\
	15  -6.36544288353527\\
	20  -6.53428789250563\\
};

\addplot [color=green,dashed,line width=0.8pt,mark=square,mark options={solid},mark size=3pt]
table[row sep=crcr]{
	-5  -1.61446643058834\\
	0	-3.27409077380377\\
	5	-4.91213225407265\\
	10  -5.96976851354680\\
	15	-6.32469492530777\\
	20  -6.44666022333306\\
};

\addplot [color=green,solid,line width=0.8pt,mark=square,mark options={solid},mark size=3pt]
table[row sep=crcr]{
	-5  -1.63976870967137\\
	0	-3.23398188414459\\
	5	-4.89336858074328\\
	10	-5.96299837687933\\
	15	-6.34829835954825\\
	20  -6.42206835529104\\
};

\addplot [color=red,dashed,line width=0.8pt,mark=diamond,mark options={solid},mark size=3pt]
table[row sep=crcr]{
	-5   3.92613375246158\\
	0   -0.398670508279070\\
	5	-3.57704721400413\\
	10	-5.19458623195698\\
	15	-5.90155609653869\\
	20  -6.13040176056085\\
};

\addplot [color=red,solid,line width=0.8pt,mark=diamond,mark options={solid},mark size=3pt]
table[row sep=crcr]{
	-5   3.90721675995952\\
	0	-0.432777422611003\\
	5	-3.57278966385417\\
	10	-5.23094230333926\\
	15	-5.92780170204334\\
	20  -6.16210295280571\\
};

\addplot [color=blue,dashed,line width=0.8pt,mark=diamond,mark options={solid},mark size=3pt]
table[row sep=crcr]{
	-5  3.55811223044383\\
	0   -0.959953161498967\\
	5	-4.54198179628130\\
	10	-6.60890099555873\\
	15  -7.48555733563607\\
	20  -7.85198628558039\\
};

\addplot [color=blue,solid,line width=0.8pt,mark=diamond,mark options={solid},mark size=3pt]
table[row sep=crcr]{
	-5  3.55209815463091\\
	0   -0.990527147920573\\
	5	-4.52724889263034\\
	10	-6.65857952365565\\
	15  -7.53670490460663\\
	20  -7.86635126824482\\
};

\addplot [color=green,dashed,line width=0.8pt,mark=diamond,mark options={solid},mark size=3pt]
table[row sep=crcr]{
	-5  3.34903058211033\\
	0   -1.30529524065451\\
	5	-4.81414172584199\\
	10	-6.87150525992877\\
	15  -7.68649077535009\\
	20  -8.02345841667490\\
};

\addplot [color=green,solid,line width=0.8pt,mark=diamond,mark options={solid},mark size=3pt]
table[row sep=crcr]{
	-5  3.30969186867809\\
	0   -1.21945761943975\\
	5	-4.84150400083305\\
	10	-6.92955327348095\\
	15  -7.72020906716728\\
	20  -7.98432102986612\\
};

\node [draw,fill=white,font=\footnotesize,anchor= south west,at={(9.5,0.9)}] {
	\setlength{\tabcolsep}{0mm}
	\renewcommand{\arraystretch}{1}
	\begin{tabular}{l}
	\ref{L24}{~LRA-LS}\\
	\ref{L21}{~LRA-LMS}\\
	\ref{L22}{~LRA-LMMSE}\\
	\end{tabular}
};

\node [draw,fill=white,font=\footnotesize,anchor= south west,at={(11,-2)}] {
	\setlength{\tabcolsep}{0mm}
	\renewcommand{\arraystretch}{1}
	\begin{tabular}{l}
	\ref{L22221}{~$|\rho|=0$}\\
	\ref{L22222}{~$|\rho|=0.75$}\\
	\end{tabular}
};

\end{axis}

\end{tikzpicture}%
    \caption{Normalized MSE comparisons of different channel estimators with adaptively estimated $\hat{\mathbf{R}}_\mathbf{h'}$.}
    \label{fig:MSE_h}
\end{figure}

Practically, $\mathbf{R}_\mathbf{h'}$ is not known at the receiver. Fig. \ref{fig:MSE_h} shows the MSE performance of the LRA channel estimators by using the proposed adaptive recursion to estimate $\mathbf{R}_\mathbf{h'}$, where $\lambda$ is set to 0.99. It can be seen that the performance remains almost the same as Fig. \ref{fig:MSE_com}, which shows that the proposed estimation of $\mathbf{R}_\mathbf{h'}$ works well under uncorrelated channel.

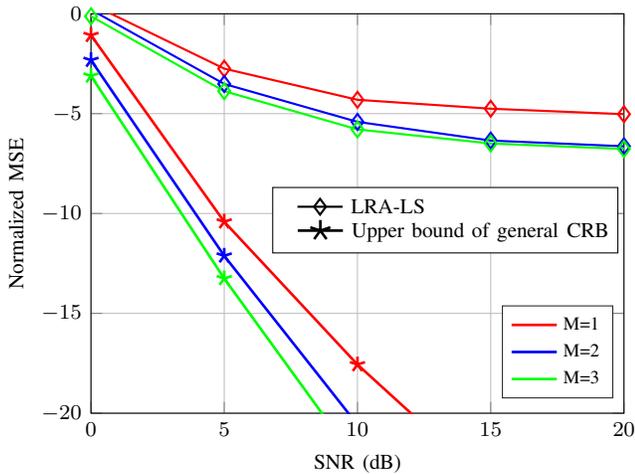
\begin{figure}[!htbp]
    \centering
    \pgfplotsset{every axis label/.append style={font=\footnotesize},
	every tick label/.append style={font=\footnotesize},
}

\begin{tikzpicture}

\begin{axis}[%
width=.8\columnwidth,
height=.6\columnwidth,
at={(0.758in,0.603in)},
scale only axis,
xmin=0,
xmax=20,
xlabel style={font=\footnotesize},
xlabel={SNR (dB)},
xtick=data,
xmajorgrids,
ymin=-20,
ymax=0,
yminorticks=true,
ylabel style={font=\footnotesize},
ylabel={Normalized MSE},
ymajorgrids,
yminorgrids,
axis background/.style={fill=white},
legend entries={M=1,
	M=2,
	M=3},
legend style={at={(0.98,0.27)},anchor=north east,legend cell align=left,align=left,draw=white!15!black,font=\scriptsize}
]

\addlegendimage{color=red,fill=gray!20,line width=1.0pt,mark size=4pt}
\addlegendimage{color=blue,fill=green!20,line width=1.0pt,mark size=4pt}
\addlegendimage{color=green,fill=gray!20,line width=1.0pt,mark size=4pt}

\addplot [color=red,solid,line width=1.0pt,mark=star,mark options={solid},mark size=3pt]
table[row sep=crcr]{%
	-5   12.5226\\
	0	-1.0743\\
	5   -10.4128\\
	10	-17.5637\\
	15  -23.6365\\
	20	-29.2527\\
};

\addplot [color=blue,solid,line width=1.0pt,mark=star,mark options={solid},mark size=3pt]
table[row sep=crcr]{%
	0	-2.3221\\
	5	-12.1189\\
	10	-20.5745\\
	15  -27.5256\\
	20	-33.6542\\
};

\addplot [color=green,solid,line width=1.0pt,mark=star,mark options={solid},mark size=3pt]
table[row sep=crcr]{%
	0	-3.1102\\
	5	-13.2551\\
	10	-22.4571\\
	15	-29.8539\\ 
	20	-36.7357\\
};

\addplot [color=red,solid,line width=0.8pt,mark=diamond,mark options={solid},mark size=3pt]
table[row sep=crcr]{%
	-5  4.84855299296409\\
	0	0.489309438443319\\
	5	-2.74532526237716\\
	10	-4.30627992964357\\
	15	-4.75301075274958\\
	20  -5.02674701281436\\
};

\addplot [color=blue,solid,line width=0.8pt,mark=diamond,mark options={solid},mark size=3pt]
table[row sep=crcr]{%
	-5  4.53405446315673\\
	0   0.203807308346261\\
	5	-3.51204495767762\\
	10  -5.41173088753114\\
	15  -6.34687547750368\\
	20  -6.63384289858440\\
};

\addplot [color=green,solid,line width=0.8pt,mark=diamond,mark options={solid},mark size=3pt]
table[row sep=crcr]{%
	-5  4.29558380915792\\
	0	-0.124114642406108\\
	5	-3.87396151017576\\
	10	-5.78961717098622\\
	15	-6.49440536834695\\
	20  -6.76777215057457\\
};

\addplot[only marks,smooth,color=black,solid,line width=1.0pt,mark=star,mark size=4pt,y filter/.code={\pgfmathparse{\pgfmathresult-0}\pgfmathresult}]
table[row sep=crcr]{%
	1 2 3 4 5 6\\
};\label{L27}

\node [draw,fill=white,font=\footnotesize,anchor= south west,at={(6.8,-12)}] {
	\setlength{\tabcolsep}{0mm}
	\renewcommand{\arraystretch}{1}
	\begin{tabular}{l}
	\ref{L24}{~LRA-LS}\\
	\ref{L27}{~Upper bound of general CRB}\\
	\end{tabular}
};
\end{axis}
\end{tikzpicture}%
    \caption{Normalized MSE comparisons of different oversampling factors for the LRA-LS channel estimator with estimated $\hat{\mathbf{R}}_\mathbf{h'}$.}
    \label{fig:MSE_h2_bias}
\end{figure}

While analyzing the general CRBs proposed in (\ref{equ_GCRB1}) and (\ref{equ_GCRB2}), instead of directly calculating the gradient of the expected value with respect to the channel vector $\frac{\partial E\{\hat{\mathbf{h'}}^{R/I}_{\text{bias}}\}}{\partial \mathbf{h'}^{R/I}}$, this gradient is numerically evaluated, since there is an adaptive estimation technique inside the channel estimator, which makes the calculation more difficult. As one example, Fig. \ref{fig:MSE_h2_bias} shows the normalized MSE performance of the LRA-LS channel estimator with estimated $\hat{\mathbf{R}}_\mathbf{h'}$ in (\ref{equ_Rh}) for estimating the first $N_r$ elements\footnote{For the sake of simplicity, only first $N_r$ elements are considered, since for the large-scale MIMO there are $N_tN_r$ elements in $\mathbf{h'}^{R}$, which will cost much time for calculating the general CRBs.} of $\mathbf{h'}^{R}$ and its corresponding numerically calculated general CRBs  under uncorrelated channels ($|\rho|=0$). More specifically, each element of the gradient vector $\frac{\partial E\{\hat{\mathbf{h'}}^{R/I}_{\text{bias}}\}}{\partial \mathbf{h'}^{R/I}}$ is calculated with the following steps:
\begin{itemize}
    \item increasing a small value $\delta$ (e.g. 0.1) in the corresponding element of $\mathbf{h'}^{R/I}$
    \item estimating the channel $\hat{\mathbf{h'}}^{R/I}_{\text{bias}}$ with different transmit symbols and noises (e.g. 1000 different realizations)
    \item calculating the mean value of all estimates $E\{\hat{\mathbf{h'}}^{R/I}_{\text{bias}}\}$, which will be divided by $\delta$.
\end{itemize}
These steps are repeated until all the elements in $\frac{\partial E\{\hat{\mathbf{h'}}^{R/I}_{\text{bias}}\}}{\partial \mathbf{h'}^{R/I}}$ are obtained.

\subsection{1-bit or b-bit ADC?}
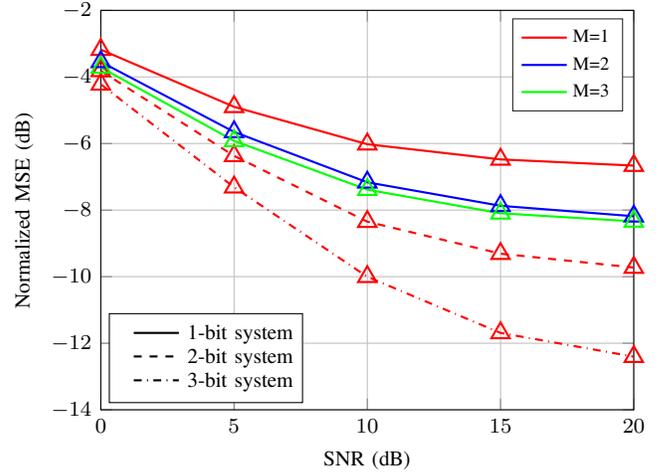
\begin{figure}[!htbp]
    \centering
    \pgfplotsset{every axis label/.append style={font=\footnotesize},
	every tick label/.append style={font=\footnotesize},
}

\begin{tikzpicture}

\begin{axis}[%
width=.8\columnwidth,
height=.6\columnwidth,
at={(0.758in,0.603in)},
scale only axis,
xmin=0,
xmax=20,
xlabel style={font=\footnotesize},
xlabel={SNR (dB)},
xtick=data,
xmajorgrids,
ymin=-14,
ymax=-2,
yminorticks=true,
ylabel style={font=\footnotesize},
ylabel={Normalized MSE (dB)},
ymajorgrids,
yminorgrids,
axis background/.style={fill=white},
legend entries={M=1,
	M=2,
	M=3},
legend style={at={(0.98,0.98)},anchor=north east,legend cell align=left,align=left,draw=white!15!black,font=\scriptsize}
]

\addlegendimage{color=red,fill=gray!20,line width=0.8pt,mark size=4pt}
\addlegendimage{color=blue,fill=green!20,line width=0.8pt,mark size=4pt}
\addlegendimage{color=green,fill=gray!20,line width=0.8pt,mark size=4pt}

\addplot [color=red,dashed,line width=0.8pt,mark=triangle,mark options={solid},mark size=4pt]
table[row sep=crcr]{%
	0	-3.81627648122915\\
	5   -6.37052615696335\\
	10	-8.34738105053021\\
	15	-9.30951679566131\\
	20  -9.72598569341110\\
};

\addplot [color=red,dashdotted,line width=0.8pt,mark=triangle,mark options={solid},mark size=4pt]
table[row sep=crcr]{%
	0   -4.22151594594495\\
	5	-7.31226924362112\\
	10	-10.0025470171241\\
	15  -11.6905885808764\\
	20  -12.4034037859200\\
};

\addplot [color=red,solid,line width=0.8pt,mark=triangle,mark options={solid},mark size=4pt]
table[row sep=crcr]{%
	0	-3.17958048378023\\
	5   -4.89703009813506\\
	10	-6.02143932322647\\
	15	-6.47733694669436\\
	20  -6.66350240619888\\
};

\addplot [color=blue,solid,line width=0.8pt,mark=triangle,mark options={solid},mark size=4pt]
table[row sep=crcr]{%
	0   -3.54632521064398\\
	5	-5.65126672243463\\
	10	-7.16522856206239\\
	15  -7.87106106194412\\
	20  -8.18454503468383\\
};

\addplot [color=green,solid,line width=0.8pt,mark=triangle,mark options={solid},mark size=4pt]
table[row sep=crcr]{%
	0	-3.71482431240083\\
	5	-5.92006064387853\\
	10	-7.38607736439999\\
	15	-8.09583441396493\\
	20	-8.33914492002643\\
};

\addplot[only marks,smooth,color=black,solid,line width=0.8pt,y filter/.code={\pgfmathparse{\pgfmathresult-0}\pgfmathresult}]
table[row sep=crcr]{%
	1 2\\
};\label{L11}

\addplot[only marks,smooth,color=black,dashed,line width=0.8pt,y filter/.code={\pgfmathparse{\pgfmathresult-0}\pgfmathresult}]
table[row sep=crcr]{%
	1 2\\
};\label{L12}

\addplot[only marks,smooth,color=black,dashdotted,line width=0.8pt,y filter/.code={\pgfmathparse{\pgfmathresult-0}\pgfmathresult}]
table[row sep=crcr]{%
	1 2\\
};\label{L13}

\node [draw,fill=white,font=\footnotesize,anchor= south west,at={(0.3,-13.8)}] {
	\setlength{\tabcolsep}{0mm}
	\renewcommand{\arraystretch}{1}
	\begin{tabular}{l}
	\ref{L11}{~1-bit system}\\
	\ref{L12}{~2-bit system}\\
	\ref{L13}{~3-bit system}\\
	\end{tabular}
};

\end{axis}

\end{tikzpicture}%
    \caption{Normalized MSE comparisons of LRA-LMMSE channel estimator with known $\mathbf{R}_\mathbf{h'}$ under uncorrelated channel ($|\rho|=0$).}
    \label{fig_morebits}
\end{figure}

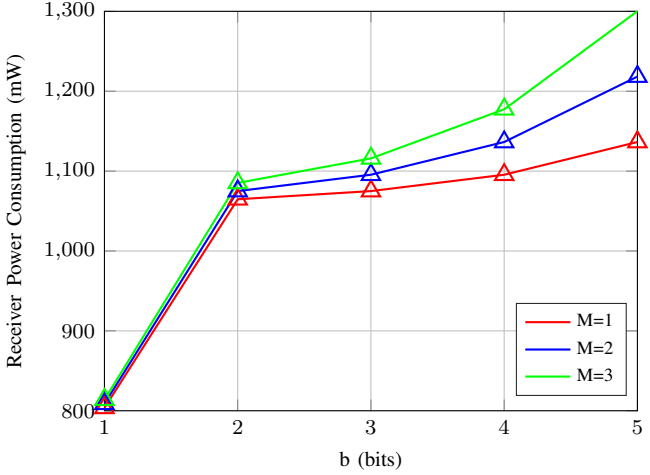
\begin{figure}[!htbp]
\centering
\pgfplotsset{every axis label/.append style={font=\footnotesize},
	every tick label/.append style={font=\footnotesize},
}

\begin{tikzpicture}

\begin{axis}[%
width=.8\columnwidth,
height=.6\columnwidth,
at={(0.758in,0.603in)},
scale only axis,
xmin=1,
xmax=5,
xlabel style={font=\footnotesize},
xlabel={b (bits)},
xtick=data,
xmajorgrids,
ymin=800,
ymax=1300,
yminorticks=true,
ylabel style={font=\footnotesize},
ylabel={Receiver Power Consumption (mW)},
ymajorgrids,
yminorgrids,
axis background/.style={fill=white},
legend entries={M=1,
	M=2,
	M=3},
legend style={at={(0.98,0.27)},anchor=north east,legend cell align=left,align=left,draw=white!15!black,font=\scriptsize}
]

\addlegendimage{color=red,solid,line width=0.8pt,mark size=4pt}
\addlegendimage{color=blue,solid,line width=0.8pt,mark size=4pt}
\addlegendimage{color=green,solid,line width=0.8pt,mark size=4pt}

\addplot [color=red,solid,line width=0.8pt,mark=triangle,mark options={solid},mark size=4pt]
table[row sep=crcr]{%
	1	803.6200\\
	2   1064.74000000000\\
	3	1074.98000000000\\
	4	1095.46000000000\\
	5   1136.42000000000\\
};

\addplot [color=blue,solid,line width=0.8pt,mark=triangle,mark options={solid},mark size=4pt]
table[row sep=crcr]{%
	1	808.7400\\
	2   1074.98000000000\\
	3	1095.46000000000\\
	4	1136.42000000000\\
	5   1218.34000000000\\
};

\addplot [color=green,solid,line width=0.8pt,mark=triangle,mark options={solid},mark size=4pt]
table[row sep=crcr]{%
	1   813.8600\\
	2	1085.22000000000\\
	3	1115.94000000000\\
	4   1177.38000000000\\
	5   1300.26000000000\\
};

\end{axis}

\end{tikzpicture}%
\caption{Receiver power consumption as a function of the quantization bits $b$.}
\label{Energy_morebits}
\end{figure}

In this subsection, the channel estimation performance of the 1-bit oversampled system is compared with the b-bit non-oversampled systems. In Fig. \ref{fig_morebits} the LRA-LMMSE channel estimator for a system with 2 or 3 bits is based on the work in \cite{Jacobsson}. It can be seen that a system with 2 or 3 bits has better MSE performance than the 1-bit system especially at high SNR.
However, the advantages of 1-bit ADCs is that they do not require automatic gain control (AGC) and linear amplifiers, and hence the corresponding radio frequency chains can be implemented with very low cost and power consumption (a few milliwatts) \cite{5288814,Li,Mo}. As one example, Fig. \ref{Energy_morebits} shows the total receiver power consumption as a function of the quantization bits $b$. The calculation of receiver power consumption is based on the work in \cite{Xiong}
\begin{equation}
\begin{aligned}
P_{\text{total}} &= P_{\text{BB}}+P_{\text{LO}}+N_r(P_{\text{LNA}}+P_{\text{H}}+2P_{\text{M}})\\&\hspace{3cm}+2N_r(cP_{\text{AGC}}+P_{\text{ADC}}),
\end{aligned}
\end{equation}
where $P_{\text{BB}}$, $P_{\text{LO}}$, $P_{\text{LNA}}$, $P_{\text{H}}$, $P_{\text{M}}$ and $P_{\text{AGC}}$ denote the power consumption in the baseband processor, local oscillator (LO), low noise amplifier (LNA), $\frac{\pi}{2}$ hybrid and LO buffer, Mixer and AGC, respectively. c is chosen as 0 for the 1-bit system and 1 for b-bit systems. The power consumption of different hardware components is given as $P_{\text{BB}}=200\text{ mW}$, $P_{\text{LO}}=22.5\text{ mW}$, $P_{\text{LNA}}=5.4\text{ mW}$, $P_{\text{H}}=3\text{ mW}$, $P_{\text{AGC}}=2\text{ mW}$ and $P_{\text{M}}=0.3\text{ mW}$. The $P_{\text{ADC}}$ is calculated as
\begin{equation}
P_{\text{ADC}} = \text{FOM}_w\times Mf_n\times 2^{b},
\end{equation}
where $\text{FOM}_w$ is 200 fJ/conversion-step at 50 MHz bandwidth and $f_n$ is 100 MHz. From the results, it can be seen that the 1-bit system consumes much less power than the 2-bit and 3-bit systems in both non-oversampled and oversampled systems. Indeed, the 1-bit oversampled systems have largely improved the estimation performance and allows the estimator to approach the performance of the 2-bit system at low SNR.

\section{Conclusion}
In this work, oversampling based low-resolution aware channel estimators have been proposed for uplink single-cell large-scale MIMO systems with 1-bit ADCs employed at the receiver. The Bussgang decomposition is used to derive linear channel estimators based on different criteria. With oversampling in such systems, it is observed that we can achieve obvious advantage compared to the non-oversampled system in terms of the normalized MSE. Moreover, the LMS adaptive technique used for channel estimation can largely reduce the computational cost and has almost the same accuracy as the LRA-LMMSE channel estimator at low SNR, which is important to ensure low computational complexity and for hardware implementation. In addition, we have also derived Bayesian and general CRBs on MSE, which give theoretical limits on the performance of the channel estimators. Furthermore, we have proposed an adaptive technique to estimate the auto-correlation of channel vector, which is important for practical use. In general, the 1-bit ADCs have the advantage of energy saving. Our proposed oversampling based channel estimation, especially the LRA-LMS estimator, increases the accuracy of estimation while maintaining low computational cost, which is important for future low cost and low latency wireless systems.

\appendices
\section{Proof of (23)}
Recall the optimization problem
\begin{equation}
\mathbf{W}_{\text{LMMSE}}=\argmin_{\mathbf{W}} E\{||\mathbf{h'}-\mathbf{W}\mathbf{y}_{\mathcal{Q}_p}||^2\}.
\end{equation}
Taking the partial derivative with respect to $\mathbf{W}^H$, we obtain
\begin{equation}
\frac{\partial E\{||\mathbf{h'}-\mathbf{W}\mathbf{y}_{\mathcal{Q}_p}||^2\}}{\partial \mathbf{W}^H}=-E\{\mathbf{h'}\mathbf{y}_{\mathcal{Q}_p}^H\}+\mathbf{W}E\{\mathbf{y}_{\mathcal{Q}_p}\mathbf{y}_{\mathcal{Q}_p}^H\}.
\label{equ_deriva}
\end{equation}
Inserting (\ref{equ_linear}) into (\ref{equ_deriva}), the LMMSE filter is
\begin{equation}
\begin{aligned}
\mathbf{W}_{\text{LMMSE}}&=E\{\mathbf{h'}\mathbf{y}_{\mathcal{Q}_p}^H\}E\{\mathbf{y}_{\mathcal{Q}_p}\mathbf{y}_{\mathcal{Q}_p}^H\}^{-1}\\&=(E\{\mathbf{h'}\mathbf{h'}^H\}\tilde{\mathbf{\Phi}}_p^H + E\{\mathbf{h'}\tilde{\mathbf{n}}_p^H\})\mathbf{C}_{\mathbf{y}_{\mathcal{Q}_p}}^{-1}.
\end{aligned}
\end{equation}
Since $\mathbf{h'}$ is uncorrelated with $\mathbf{n}_p$ and $\mathbf{n}_q$ \cite{Li}, we have
\begin{equation}
E\{\mathbf{h'}\tilde{\mathbf{n}}_p^H\}=E\{\mathbf{h'}(\mathbf{A}_p\mathbf{n}_p+\mathbf{n}_q)^H\}=\mathbf{0}.
\end{equation}
The resulting LRA-LMMSE channel estimator is
\begin{equation}
\hat{\mathbf{h'}}_{\text{LRA-LMMSE}}=\mathbf{R}_\mathbf{h'}\tilde{\mathbf{\Phi}}_p^H\mathbf{C}_{\mathbf{y}_{\mathcal{Q}_p}}^{-1}\mathbf{y}_{\mathcal{Q}_p}.
\end{equation}

\section{Proof of (29)}
Defining $\mathbf{\epsilon}(n)=\bar{\mathbf{h'}}^{n_r}(n)-\mathbf{h'}^{n_r}$ and inserting it into (\ref{equ_recursion}), we obtain
\begin{equation}
\begin{aligned}
\mathbf{\epsilon}(n+1)&=\mathbf{\epsilon}(n)+\mu \tilde{\mathbf{\Phi}}^{n_r}_p(n)^H(\mathbf{y}^{n_r}_{\mathcal{Q}_p}(n)-\tilde{\mathbf{\Phi}}^{n_r}_p(n)\bar{\mathbf{h'}}^{n_r}(n))\\&=\mathbf{\epsilon}(n)+\mu \tilde{\mathbf{\Phi}}^{n_r}_p(n)^H\mathbf{y}^{n_r}_{\mathcal{Q}_p}(n)\\&\hspace{1cm}-\mu \tilde{\mathbf{\Phi}}^{n_r}_p(n)^H\tilde{\mathbf{\Phi}}^{n_r}_p(n)(\mathbf{\epsilon}(n)+\mathbf{h'}^{n_r})\\&=(\mathbf{I}-\mu \tilde{\mathbf{\Phi}}^{n_r}_p(n)^H\tilde{\mathbf{\Phi}}^{n_r}_p(n))\mathbf{\epsilon}(n)\\&\hspace{1cm}+\mu \tilde{\mathbf{\Phi}}^{n_r}_p(n)^H(\mathbf{y}^{n_r}_{\mathcal{Q}_p}(n)-\tilde{\mathbf{\Phi}}^{n_r}_p(n)\mathbf{h'}^{n_r}).
\end{aligned}
\end{equation}
Taking the expected value from $\mathbf{\epsilon}(n+1)$, we have
\begin{equation}
E\{\mathbf{\epsilon}(n+1)\}=(\mathbf{I}-\mu E\{\tilde{\mathbf{\Phi}}^{n_r}_p(n)^H\tilde{\mathbf{\Phi}}^{n_r}_p(n)\})E\{\mathbf{\epsilon}(n)\}.
\label{equ_expec}
\end{equation}
With the eigenvalue decomposition $E\{\tilde{\mathbf{\Phi}}^{n_r}_p(n)^H\tilde{\mathbf{\Phi}}^{n_r}_p(n)\}=\mathbf{Q}\mathbf{\Gamma}\mathbf{Q}^H$, (\ref{equ_expec}) can be written as
\begin{equation}
\begin{aligned}
\mathbf{Q}^HE\{\mathbf{\epsilon}(n+1)\}&=\mathbf{Q}^H(\mathbf{I}-\mu \mathbf{Q}\mathbf{\Gamma}\mathbf{Q}^H)E\{\mathbf{\epsilon}(n)\}\\&=(\mathbf{I}-\mu \mathbf{\Gamma})\mathbf{Q}^HE\{\mathbf{\epsilon}(n)\},
\end{aligned}
\label{equ_QQ}
\end{equation}
where $\mathbf{Q}$ is an unitary matrix and $\mathbf{\Gamma}$ is a diagonal matrix, whose diagonal entries are the eigenvalues of $E\{\tilde{\mathbf{\Phi}}^{n_r}_p(n)^H\tilde{\mathbf{\Phi}}^{n_r}_p(n)\}$. With $\mathbf{u}(n)=\mathbf{Q}^HE\{\mathbf{\epsilon}(n)\}$, (\ref{equ_QQ}) is then
\begin{equation}
\mathbf{u}(n+1)=(\mathbf{I}-\mu \mathbf{\Gamma})\mathbf{u}(n).
\end{equation}
Decoupling the matrix form into individual elements we get
\begin{equation}
\begin{aligned}
u_{n_t}(n+1)&=(1-\mu \gamma_{n_t})u_{n_t}(n)\\&=(1-\mu \gamma_{n_t})^{\tau-l_{\text{win}}+1} u_{n_t}(1),\hspace{1em} n_t=1,\ldots,N_t.
\end{aligned}
\end{equation}
In order for the LRA-LMS to converge, we must have
\begin{equation}
|1-\mu \gamma_{n_t}|<1.
\end{equation}
The stability condition is then given by
\begin{equation}
0<\mu<\frac{2}{\gamma_{\max}},
\label{equ_stablitycon}
\end{equation}
where $\gamma_{\max}$ is the largest eigenvalue of $E\{\tilde{\mathbf{\Phi}}^{n_r}_p(n)^H\tilde{\mathbf{\Phi}}^{n_r}_p(n)\}$.

\section{Proof of Lemma 1}
The biasness of the adaptive estimator $\hat{\mathbf{R}}_\mathbf{h'}$ is firstly examined. The expected value of $\hat{\mathbf{h'}}(n)$ in (\ref{equ_esth}) is
\begin{equation}
E\{\hat{\mathbf{h'}}(n)\} =  E\{(\mathbf{x'}_p^T(n)\otimes\mathbf{I}_{N_r}\otimes\mathbf{Z'u})^+\mathbf{y}_\mathcal{Q}(n)\}.
\label{equ_expechn}
\end{equation}
From the Bussgang theorem (\ref{equ_sysmodeln}) can be decomposed as
\begin{equation}
\resizebox{\columnwidth}{!}{$\displaystyle
\begin{aligned}
\mathbf{y}_\mathcal{Q}(n)&=\mathcal{Q}\left((\mathbf{x'}_p^T(n)\otimes\mathbf{I}_{N_r}\otimes\mathbf{Z'u})\mathbf{h'}+\mathbf{n}(n)\right)\\&=\mathbf{A'}_p(n)((\mathbf{x'}_p^T(n)\otimes\mathbf{I}_{N_r}\otimes\mathbf{Z'u})\mathbf{h'}+\mathbf{n}(n))+\mathbf{n}_q(n),
\end{aligned}$}
\label{equ_yQ}
\end{equation}
where $\mathbf{A'}_p(n)$ is the linear operator and $\mathbf{n}_q(n)$ is the statistically equivalent quantizer noise. Substituting (\ref{equ_yQ}) into (\ref{equ_expechn}) and with $\mathbf{\Phi'}(n)=(\mathbf{x}^T(n)\otimes\mathbf{I}_{N_r}\otimes\mathbf{Z'u})$, we obtain
\begin{equation}
\resizebox{\columnwidth}{!}{$\displaystyle
\begin{aligned}
E\{\hat{\mathbf{h'}}(n)\} &=  E\{\mathbf{\Phi'}(n)^+(\mathbf{A'}_p(n)(\mathbf{\Phi'}(n)\mathbf{h'}+\mathbf{n}(n))+\mathbf{n}_q(n))\}\\&=E\{\mathbf{\Phi'}(n)^+\mathbf{A'}_p(n)\mathbf{\Phi'}(n)\mathbf{h'}\}\\&+E\{\mathbf{\Phi'}(n)^+\mathbf{A'}_p(n)\mathbf{n}(n)\}+E\{\mathbf{\Phi'}(n)^+\mathbf{n}_q(n)\}.
\end{aligned}$}
\label{equ_expechnprime}
\end{equation}
Since $\mathbf{\Phi'}(n)$ and $\mathbf{n}(n)$ are uncorrelated and $E\{\mathbf{n}(n)\}=\mathbf{0}$, we have
\begin{equation}
E\{\mathbf{\Phi'}(n)^+\mathbf{A'}_p(n)\mathbf{n}(n)\} = \mathbf{0}.
\end{equation}
Similarly,
\begin{equation}
E\{\mathbf{\Phi'}(n)^+\mathbf{n}_q(n)\} = \mathbf{0}.
\end{equation}
Equation (\ref{equ_expechnprime}) can be further simplified as
\begin{equation}
E\{\hat{\mathbf{h'}}(n)\} =  E\{\mathbf{\Phi'}(n)^+\mathbf{A'}_p(n)\mathbf{\Phi'}(n)\}\mathbf{h'}.
\label{equ_hequ}
\end{equation}
The matrix $\mathbf{A'}_p(n)$ depends on $\mathbf{R}_\mathbf{h'}$ such that the expectation in (\ref{equ_hequ}) can be different from the identity matrix especially for channels without normalization, which verifies that (\ref{equ_esth}) has an unknown bias \cite{Trees}. With the analysis above, it is concluded that the adaptive estimator $\hat{\mathbf{R}}_\mathbf{h'}$ is also biased, which shows that the estimation procedures together with the proposed LRA channel estimators are biased.

\bibliographystyle{IEEEtran}
\bibliography{ref}

\end{document}